\documentclass[aps,prc,twocolumn,showpacs,amsmath,amssymb,nofootinbib]{revtex4-1}
\usepackage{blindtext}
\usepackage{graphicx}
\usepackage{booktabs}
\usepackage{color}
\usepackage{times}
\usepackage{bm}
\usepackage{multirow}
\usepackage{float}
\usepackage{url}
\usepackage{natbib}
\usepackage{tensor}
\usepackage{bbm}
\usepackage[colorlinks=true,citecolor=blue,urlcolor=blue,linkcolor=red]{hyperref}
\usepackage[normalem]{ulem}
\usepackage{verbatim}
\usepackage{lipsum} 
\usepackage{tikz,xcolor,hyperref}
\usepackage{ifxetex,ifluatex}
\usepackage{lipsum} 
\usepackage{tikz,xcolor,hyperref}

\newcommand{\apjl}{Astrophys. J. Lett.}

\newcommand{\aap}{Astron. Astrophys.}
\newcommand{\araa}{Annual Review of Astronomy \& Astrophysics}
\newcommand{\mnras}{Mon. Not. R. Astron. Soc.}
\newcommand{\ssr}{Space Science Reviews}
\newcommand{\nphysa}{Nuclear Physics A}

\definecolor{lime}{HTML}{A6CE39}
\DeclareRobustCommand{\orcidicon}{%
	\begin{tikzpicture}
	\draw[lime, fill=lime] (0,0) 
	circle [radius=0.17] 
	node[white] {{\fontfamily{qag}\selectfont \tiny ID}};
	\draw[white, fill=white] (-0.0625,0.095) 
	circle [radius=0.008];
	\end{tikzpicture}
	\hspace{-2mm}
}
\foreach \x in {A, ..., Z}{
	\expandafter\xdef\csname orcid\x\endcsname{\noexpand\href{https://orcid.org/\csname orcidauthor\x\endcsname}{\noexpand\orcidicon}}
}

\begin{document}

\title{Unified QMF equation of state for neutron star matter: Static and dynamic properties}
\author{Zhonghao Tu{\orcidD{}}$^{1}$}
\author{Xiangdong Sun$^{1}$}
\author{Shuochong Han$^{1}$}
\author{Zhiqiang Miao{\orcidA{}}$^{2}$}
\author{Ang Li{\orcidB{}}$^{1}$}
\affiliation{\it $^{1}$Department of Astronomy, Xiamen University, Xiamen 361005, China; \textit{liang@xmu.edu.cn}; \\
$^{2}$ Tsung-Dao Lee Institute, Shanghai Jiao Tong University, Shanghai 201210, China
}

\begin{abstract}
We construct a set of unified equations of state based on the quark mean-field (QMF) model, calibrated to different values of nuclear symmetry energy slope at the saturation density ($L_0$), with the aim of exploring both the static properties and dynamical behavior of neutron stars (NSs), and building a coherent picture of their internal structure.
We assess the performance of these QMF models in describing the mass-radius relation, the cooling evolution of isolated NSs and X-ray transients, and the instabilities (e.g., the r-mode).
In comparison to relativistic mean-field (RMF) models formulated at the hadronic level, the QMF model predicts heavier nuclear clusters and larger Wigner-Seitz cell sizes in the NS crust, while the density of the free neutron gas remains largely similar between the two approaches.
For the cooling of isolated NSs, the thermal evolution is found to be insensitive to both the many-body model and the symmetry energy slope in the absence of the direct Urca (dUrca) process. However, when rapid cooling via the dUrca process is allowed, in the case of large $L_0$ values (e.g., $L_0 \gtrsim 80$ MeV) in our study, the QMF model predicts a longer thermal relaxation time. 
Both the QMF and RMF models can reproduce cooling curves consistent with observations of X-ray transients (e.g., KS 1731--260) during their crustal cooling phase, although stellar parameters show slight variations depending on the model and symmetry energy slope.
Within our unified framework, a larger $L_0$ value generally results in a wider instability window, while increasing the stellar mass tends to suppress the instability window.
We also provide simple power-law parameterizations that quantify the dependence of bulk and shear viscosities on the symmetry energy slope for nuclear matter at saturation density.
\end{abstract}

\maketitle

\section{Introduction}
The equation of state (EoS) of dense and highly isospin asymmetric matter 
remains a central challenge in nuclear physics and astrophysics, as it governs not only the global structure of neutron stars (NSs) but also their thermal and rotational evolution \cite{2015ARNPS..65..303G,2017RvMP...89a5007O,2016RvMP...88b1001W,2004ARA&A..42..169Y,1995SSRv...74..427H,2019RAA....19...72Y}.
Advances in experiments and observations, such as laboratory measurements of finite nuclei \cite{2022PhRvL.129d2501A,2021PhRvL.126q2502A}, constraints on symmetric nuclear matter up to twice the nuclear saturation density from heavy-ion collisions \cite{2002Sci...298.1592D,2006PrPNP..56....1F}, and observations of NSs using  
this-generation observational facilities \cite{2017PhRvL.119p1101A,2024ApJ...961...62V,2024ApJ...974..294S} have provided a solid foundation for developing phenomenological many-body approaches that describe nuclear matter over a wide range of densities and isospin asymmetries. 

Among these models, the relativistic mean-field (RMF) model~\citep{1974AnPhy..83..491W,2011PrPNP..66..519N} offers a variable framework for describing nuclear matter and NS matter by incorporating accumulated experimental and observational data~\cite[see recent works in e.g., ][]{2025arXiv250323028P}. In the RMF model, nucleons are treated as point-like particles interacting via meson exchange. 
The quark mean-field (QMF) model~\citep{2000PhRvC..61d5205S,1998PhRvC..58.3749T} incorporates quark-level degrees of freedom while retaining a realistic description of hadronic matter. 
The QMF model describes interactions through direct coupling of quarks to mesons and gluons, leading to the evolution of medium nucleon mass with quark mass corrections. Thus, the most significant difference between the two models lies in the density dependence of the in-medium nucleon mass and radius~\cite{2018PhRvC..97c5805Z,2019PhRvC..99b5804Z}.
The QMF model has been successfully applied to describe finite nuclei, nuclear matter, and NS matter within a single theoretical framework~\cite[see recent progress in e.g., ][]{2018ApJ...862...98Z,2019PhRvC..99b5804Z,2020ApJ...904..103M,2022JPhG...49b5104H,2024PhRvC.110e5201M,2025arXiv250403498P,2025arXiv250418874C}, making it particularly suitable for a comprehensive study of NS physics.

The exploration of dense matter properties within NSs requires a theoretical framework capable of describing not only their static properties — such as their mass-radius relation, internal composition, and maximum mass limits — but also their dynamic evolution, including thermal cooling and rotational instabilities.
A key challenge in this context is the construction of a unified EoS.
The global properties of NSs are sensitive to the stiffness of both the crust and core EoSs \cite{2023ApJ...957...41L,2024JPhG...51f5203S,2016PhRvC..94c5804F,2021NuPhA101022189R}. The cooling behavior of isolated NSs and the crustal thermal relaxation in X-ray transients depend on the internal composition and EoS, with the former primarily governed by the core \cite{2015SSRv..191..239P,2019MNRAS.487.2639R} and the latter more influenced by the crust \cite{2022ApJ...925..205S,2023MNRAS.522.4830P}. Hot, rapidly rotating NSs are subject to r-mode instabilities, which can, however, be suppressed by viscous damping \cite{PhysRevD.58.084020}. This damping arises from bulk and shear viscosities, which are themselves sensitive to the properties of both the crust and the core \cite{2021ApJ...910...62Z,2015PhRvC..91c5804M}. 

Nonunified EoSs are obtained by matching the EoSs of three segments, e.g., the outer crust, inner crust, and core of NSs. The EoSs in different segments may be calculated by using different nuclear forces and nuclear many-body approaches, resulting in these segments having different theoretical bases. Therefore, nonunified EoSs introduce uncertainty for the description of the global and dynamical properties in a consistent manner in the interior~\cite{2021PhRvC.104a5801S,2016PhRvC..94c5804F}.
A unified EoS requires that the EoSs for three segments are calculated using the same nuclear many-body approach based on the same nuclear force, with self-consistent compositions for each segment~\cite{2013A&A...559A.128F}. A unified EoS reduces the additional degrees of freedom artificially introduced in describing NS properties, allowing them to depend more directly on the nuclear force and the many-body approach~\cite{2015A&A...584A.103S,2025PhRvC.111a5805C}. 
This strengthens the constraints on the nuclear theory from observations. 
Indeed, RMF/QMF unified EoSs have been adopted for the model calculation and Bayesian inference of nuclear matter properties from multi-messenger NS observations~\cite{2021ApJ...913...27L,2022MNRAS.515.5071M,2023PhRvC.108b5809Z}. 

In this work, we conduct a systematic study of unified EoSs for NS matter by using the QMF, in comparison with those of RMF. 
To determine the meson couplings in both models, we specifically  fit them to reproduce the empirical values of the symmetry energy slope while keeping other nuclear saturation properties fixed. 
The calculations of the crust EoS follow our previous work \cite{2025ApJ...984..200T}, with the QMF calculation mimicking the procedure for the RMF calculation, though with significantly greater technical complexity.
Using these unified EoSs, we will assess the performance of the QMF model in describing dense matter and static mass-radius relations.
We also compare in detail the resulting NS compositions and related dynamic properties of NSs predicted by the QMF and RMF models.

This paper is organized as follows. The theoretical frameworks is given in Sec. \ref{sec:theory}. Sec.~\ref{sec:theory_QMFcore}--\ref{sec:theory_QMFcrust} introduce that how to construct the unified EoS, an approximately unified EoS is given in Sec. \ref{sec:theory_Zdunik} for comparison. Sec. \ref{sec:theory_MR}--\ref{sec:theory_viscosity} is dedicated to reviewing the methodology for calculating the global properties, thermal evolution, and r-mode instability of NSs. The results and discussions are shown in Sec.~\ref{sec:result}. From Sec. \ref{sec:result_EOS}--Sec. \ref{sec:result_viscosity}, we demonstrate the dependence of the stellar composition, the cooling curves, the viscosity, and the instability of the r-mode on the uncertain nuclear symmetry energy as well as the stellar mass.
Finally, a brief summary is given in Sec.~\ref{sec:summary}.

\section{Theoretical framework}\label{sec:theory}

\subsection{The QMF model for uniform matter in NS core}\label{sec:theory_QMFcore}

The QMF model starts from the Dirac equation, which describes constituent quarks confined within a nucleon by a phenomenological confinement potential $U(r)$. Further details are available in our previous review~\cite{2020JHEAp..28...19L}; below, we present the necessary formulas.
The Dirac equation is written as
\begin{eqnarray}\label{eq:DECQ}
&& [\gamma^{0}(\epsilon_{q}-g_{\omega q}\omega-\tau_{3q}g_{\rho q}\rho) \nonumber \\
&&
-\vec{\gamma}\cdot\vec{p} -(m_{q}-g_{\sigma q}\sigma)-U(r)]\psi_{q}(\vec{r})=0\ ,
\end{eqnarray}
where $U(r)$ is taken as a harmonic oscillator potential, allowing the Dirac equation to be solved analytically. Here $\psi_{q}(\vec{r})$ is the quark field, $\sigma$, $\omega$, and $\rho$ are the classical meson fields. The coupling constants between quarks and mesons are denoted by $g_{\sigma q}$, $g_{\omega q}$, and $g_{\rho q}$, respectively. $m_q$ and $\epsilon_{q}$ are the mass and energy of quarks, respectively.
The constitute quark mass $m_q$ is taken to be 300 MeV, and $\epsilon_{q}$ is the quark energy. $\tau_{3q}$ represents the third component of the quark isospin matrix.
The ground-state energy solution from Eq.~(\ref{eq:DECQ}) satisfies
\begin{eqnarray}
(\mathop{\epsilon'_q-m'_q})\sqrt{\frac{\lambda_q}{a}}=3\ ,
\end{eqnarray}
where $\lambda_q=\epsilon_q^\ast+m_q^\ast,\ \mathop{\epsilon'_q}=\epsilon_q^\ast-V_0/2\ ,\ \mathop{m'_q}=m_q^\ast+V_0/2$. 
The effective single-quark energy and mass are given by
$\epsilon_q^*=\epsilon_{q}-g_{\omega q}\omega-\tau_{3q}g_{\rho q}\rho$ and the effective quark mass by $m_q^\ast = m_q-g_{\sigma q}\sigma$. 
The zeroth-order energy of the nucleon core is obtained as $E_N^0=\sum_q\epsilon_q^\ast$.
Corrections for the center-of-mass motion ($\epsilon_{\rm c.m.}$), quark–pion coupling ($\delta M_N^\pi$), and one-gluon exchange ($(\Delta E_N)_g$) are included to determine the in-medium nucleon mass (see details in e.g., Ref.~\cite{2018ApJ...862...98Z}):
\begin{eqnarray}
M^\ast_N=E^{0}_N-\epsilon_{\rm c.m.}+\delta M_N^\pi+(\Delta E_N)_g\ .
\end{eqnarray}
The nucleon radius is given by
\begin{eqnarray}
\langle r_N^2\rangle = \frac{\mathop{11\epsilon'_q + m'_q}}{\mathop{(3\epsilon'_q + m'_q)(\epsilon'^2_q-m'^2_q)}}\ .
\end{eqnarray}
The potential parameters $a$ and $V_0$ are determined by reproducing the free-space nucleon mass and radius, namely $M_N = 939$ MeV and $r_N = 0.87$ fm. 

To describe nuclear matter, we consider a Lagrangian with $\sigma$, $\omega$, and $\rho$ meson exchanges:
\begin{eqnarray}
\mathcal{L}& = & \bar{\psi}\left(i\gamma_\mu \partial^\mu - M_N^\ast - g_{\omega N}\omega\gamma^0 - g_{\rho N}\rho\tau_{3}\gamma^0\right)\psi_N  \nonumber \\
&&  - \frac{1}{2}m_\sigma^2 \sigma^2 - \frac{1}{3} g_2\sigma^3 - \frac{1}{4}g_3\sigma^4 \nonumber \\
&&  + \frac{1}{2}m_\omega^2\omega^2 + \frac{1}{2}m_\rho^2\rho^2+ \frac{1}{2}g_{\rho N}^2\rho^2 \Lambda_v g_{\omega N}^2\omega^2\ ,
\label{eq:L}
\end{eqnarray}
where $g_{\omega N}$ and $g_{\rho N}$ are the nucleon couplings to the $\omega$ and $\rho$ mesons. Based on the quark counting rule, we take $g_{\omega N} = 3g_{\omega q}$ and $g_{\rho N} = g_{\rho q}$. $\Lambda_v$ represents the $\omega$–$\rho$ meson cross-coupling constant. $\psi_N$ is the nucleon field, and $\tau_3$ is the third component of the isospin operator. The effective nucleon mass $M_N^*$ is obtained from the calculation of confined quarks and depends on the $\sigma$ field via the parameter $g_{\sigma q}$. The meson masses are set as: $m_\sigma = 510$ MeV, $m_\omega = 783$ MeV, and $m_\rho = 770$ MeV.

In uniform nuclear matter, the equations of motion for nucleons and mesons read as the variation of the Lagrangian:
\begin{eqnarray}
 (i\gamma^{\mu}\partial_{\mu} - M_N^* - g_{\omega N}\omega\gamma^0 - &g_{\rho N}\rho\tau_3\gamma^0)\psi_N=0\ ,\nonumber\\
 m_{\sigma}^2\sigma + g_2\sigma^2 + g_3\sigma^3 &= -\frac{\partial M_N^*}{\partial \sigma}\left\langle{\bar\psi_N}\psi_N\right\rangle\ , \nonumber \\
 m_{\omega}^2\omega + \Lambda_v g_{\omega N}^2g_{\rho N}^2\omega\rho^2&= g_{\omega N}\left\langle{\bar\psi_N}\gamma^0\psi_N\right\rangle\ , \nonumber \\
 m_{\rho}^2\rho + \Lambda_v g_{\omega N}^2g_{\rho N}^2\omega^2\rho&= g_{\rho N}\left\langle{\bar\psi_N}\tau_3\gamma^0\psi_N\right\rangle\ .
\label{eq: Core motion}
\end{eqnarray}
The energy density and pressure can be derived from the energy-momentum tensor using the Lagrangian and meson field equations:
\begin{eqnarray}
{\varepsilon}_{\rm{QMF},core} &=
& \frac{1}{\pi^2}\sum_{i=n,p}\int_{0}^{k_{F_i}}k^2\sqrt{k^2+M_N^{*2}}\mathrm{d}k\nonumber\\ 
&&+ \frac{1}{2}m_{\sigma}^2\sigma^2 + \frac{1}{3}g_2\sigma^3 +\frac{1}{4}g_3\sigma^4 \nonumber\\
&&+ \frac{1}{2}m_{\omega}^2\omega^2 + \frac{1}{2}m_{\rho}^2\rho^2 \nonumber\\
&&+\frac{3}{2}\Lambda_v g_{\omega N}^2 g_{\rho N}^2 \rho^2  \omega^2, \label{eq:ecore}\\
P_{\rm{QMF},core} &=
& \frac{1}{3\pi^2}\sum_{i=n,p}\int_{0}^{k_{F_i}}\frac{k^4}{\sqrt{k^2+M_N^{*2}}}dk  \nonumber\\ 
&&- \frac{1}{2}m_{\sigma}^2\sigma^2 - \frac{1}{3}g_2\sigma^3 -\frac{1}{4}g_3\sigma^4 \nonumber\\
&&+ \frac{1}{2}m_{\omega}^2\omega^2 + \frac{1}{2}m_{\rho}^2\rho^2 \nonumber\\
&&+\frac{3}{2}\Lambda_v g_{\omega N}^2 g_{\rho N}^2 \rho^2  \omega^2.
\label{eq:pcore}
\end{eqnarray}
Here, the kinetic terms in $\varepsilon_{\rm QMF,core}$ and $P_{\rm QMF,core}$ are integrated from zero to the Fermi momentum $k_F$. 

For $\beta$-equilibrated matter in NS cores, charge neutrality and $\beta$ equilibrium conditions are imposed to determine the particle fractions. 
The kinetic contributions from electrons and muons must also be included in Eqs.~(\ref{eq:ecore}) and (\ref{eq:pcore}). 

The Lagrangian Eq.~(\ref{eq:L}) contains six free parameters: $g_{\sigma q}$, $g_{\omega q}$, $g_{\rho q}$, $g_2$, $g_3$, and $\Lambda_v$. These are determined by fitting to the saturation properties of nuclear matter, as summarized in Table~\ref{tab:sat}. In our study, the symmetry energy slope $L_0$, which significantly influences NS structure and dynamics, is varied within the range of 40–80 MeV, while other saturation properties are kept fixed. The same procedure is applied in the RMF model, where the meson–quark couplings are replaced by meson–nucleon couplings. 

\begin{table}
\caption{Saturation properties used in this study to fit the meson coupling parameters include the saturation density $\rho_0$ (in fm$^{-3}$), along with the corresponding values at saturation for the binding energy per nucleon $E/A$ (in MeV), incompressibility $K_0$ (in MeV), symmetry energy $J_0$ (in MeV), symmetry energy slope $L_0$ (in MeV), and the ratio of the effective mass to the free nucleon mass $M_N^\ast/M_N$.
}
\setlength{\tabcolsep}{6pt}
\renewcommand{\arraystretch}{1.2}
\begin{ruledtabular}
\begin{tabular*}{\hsize}{@{}@{\extracolsep{\fill}}cccccc@{}}
$\rho_0$ & $E/A$  & $K_0$ & $J_0$& $L_0$ 
& $M_N^\ast/M_N$ \\ 
$({\rm fm}^{-3})$ & (MeV) & (MeV) & (MeV) & (MeV) \\
\hline
$0.16$ & $-16$ & $240$ & $31$ & $40/60/80$ & $0.77$ 
\label{tab:sat}
\end{tabular*}
\end{ruledtabular}
\end{table}

\subsection{Unified description of NS crust from solving Klein–Gordon equation}\label{sec:theory_QMFcrust}

In the QMF model, the coupling constant between nucleons and the $\sigma$ meson, $g_{\sigma N}=\partial M_{N}^*/\partial\sigma$, is determined self-consistently from the quark level. We derive a unified description of the crust starting from the Lagrangian (\ref{eq:L}), while keeping in mind that our approach does not involve direct derivation from the quark level, which limits the degree of theoretical unification. At a given baryon number density $\rho_B$ within a Wigner-Seitz (WS) cell of radius $R_{W}$, the meson fields and fermion density distributions can be obtained by iteratively solving the Klein-Gordon equations under the constraints of $\beta$ equilibrium and global charge neutrality:
\begin{eqnarray}
 (-\nabla^2 +m_{\sigma}^2)\sigma + g_2\sigma^2 + g_3\sigma^3 + \frac{\partial M_N^*}{\partial \sigma}\rho_s &=0 \ ,\qquad \label{eq:motion1} \\
 (-\nabla^2 +m_{\omega}^2)\omega + \Lambda_v g_{\omega N}^2g_{\rho N}^2\omega\rho^2- g_{\omega N}\rho_B &=0 \ , \qquad \label{eq:motion2} \\
 (-\nabla^2 +m_{\rho}^2)\rho + \Lambda_v g_{\omega N}^2g_{\rho N}^2\omega^2\rho- \sum_{i=n,p}g_{\omega N}\tau_i\rho_i &=0 \ ,  \qquad \label{eq:motion3}\\
 \nabla^2A + e(\rho_p - \rho_e) &=0 \ ,\qquad
\label{eq:motion4}
\end{eqnarray}
with reflective boundary conditions $\frac{d\phi(r)}{dr}|_{r=0,R_{\mathrm{W}}}=0$, $\phi=\sigma,~\omega,~\phi$, and $A$. Here, $\rho_{s}$ is the scalar density, while $\rho_i$ denotes the vector density of particle species $i$, $i=n,~p$, and $e$. The Thomas-Fermi approximation (TFA) is adopted, treating fermion wave functions as plane waves. To enhance computational efficiency, Eqs. (\ref{eq:motion1})–(\ref{eq:motion4}) are solved using a fast cosine transform.
To determine the ground state of the WS cell, the resulting matter distribution must satisfy the constancy of chemical potentials at the end of the iteration. For more detailed discussions, please refer to our previous work \citep{2025ApJ...984..200T}. 
From Eqs.~(\ref{eq:motion1})--(\ref{eq:motion4}), we can see that the photon $A$ is coupled to the nucleon; the effects of charge screening are self-consistently included in this work. 

The computational treatment differs slightly between the outer and inner crusts. In the inner crust, nuclei are generally spherical; however, near the crust-core boundary, non-spherical structures (known as nuclear pasta) may emerge due to the competition between nuclear surface tension and Coulomb forces. To simplify the modeling, we consider five representative pasta geometries: droplets, rods, slabs, tubes, and bubbles. By applying the spherical and cylindrical WS approximations, the differential equations for meson fields are reduced to one-dimensional form in the respective geometries:
\begin{eqnarray}
1\mathrm{D}: \nabla^2\phi(\boldsymbol{r})& =& \frac{d^2\phi(r)}{dr^2}  \ , \label{eq:filed1}\\
2\mathrm{D}: \nabla^2\phi(\boldsymbol{r})& =& \frac{d^2\phi(r)}{dr^2} + \frac{1}{r}\frac{d\phi(r)}{dr} \ ,\label{eq:filed2}\\
3\mathrm{D}: \nabla^2\phi(\boldsymbol{r})& =& \frac{d^2\phi(r)}{dr^2} + \frac{2}{r}\frac{d\phi(r)}{dr} \ ,
\label{eq:filed3}
\end{eqnarray}
which significantly reduces computational cost. 
For each geometry, the optimal matter configuration is found by varying the WS cell size to minimize the total energy:
\begin{equation}
E_{\rm{cell}}=\int_{\rm{cell}}{\varepsilon}_{\rm{QMF,crust}}(\boldsymbol{r})\mathrm{d} \boldsymbol{r} \ ,
\label{eq:TFe}
\end{equation}
where the energy density ${\varepsilon}_{\rm{QMF,crust}}({\boldsymbol{r}})$ is given by:
\begin{eqnarray}
{\varepsilon}_{\rm{QMF},crust} &=
& \varepsilon_{\mathrm{kin}}  \nonumber + \frac{1}{2}(\nabla \sigma)^2 + \frac{1}{2}m_{\sigma}^2\sigma^2 + \frac{1}{3}g_2\sigma^3 +\frac{1}{4}g_3\sigma^4 \nonumber\\
&&+ \frac{1}{2}(\nabla \omega)^2 + \frac{1}{2}m_{\omega}^2\omega^2 \nonumber\\
&&+ \frac{1}{2}(\nabla \rho)^2 + \frac{1}{2}m_{\rho}^2\rho^2 \nonumber\\
&&+\frac{3}{2}\Lambda_v g_{\omega N}^2 g_{\rho N}^2 \rho^2  \omega^2 + \frac{1}{2}(\nabla A)^2 \ ,
\label{eq:Tensor}
\end{eqnarray}
with the kinetic energy density $\varepsilon_{\mathrm{kin}}$ consistent with Eq. (\ref{eq:ecore}).

The pressure is evaluated using the thermodynamic relation:
\begin{eqnarray}
P_{\mathrm{QMF,crust}}=\sum_{i=n,p,e} \mu_{i} \rho_{i} - \varepsilon \ ,
\label{eq:Pre}
\end{eqnarray}
where $\mu_i$ are the chemical potentials of nucleons and electrons, and $\varepsilon=\rho_BE_{\rm{cell}}/A_{\mathrm{cell}}$  is the energy density of the WS cell, with $A_{\mathrm{cell}}$ being the total number of nucleons in the cell.
Among the five nuclear pasta configurations, the one with the lowest energy per nucleon is identified as the ground state. The crust-core transition density $\rho_{\rm{cc}}$ is then determined as the density at which the energy of uniform matter becomes lower than that of the most stable non-uniform phase.

From the nucleon density distributions, the characteristic properties of typical nuclear pasta structures can be extracted. The pasta size $R_{\rm{d}}$ is estimated as:
\begin{equation}
R_{\rm{d}}=\left\{
\begin{array}{ll}
R_\text{w}\left(\frac{\langle \rho_p \rangle}{\langle \rho^2_p \rangle}\right)^{1/D}    &\hspace{0.5cm} \textrm{droplet-like}\ ,  \\
R_\text{w}\left((1-\frac{\langle \rho_p \rangle^2}{\langle \rho^2_p \rangle}\right)^{1/D}   &\hspace{0.5cm} \textrm{bubble-like} \ ,        
\end{array}
\right.
\label{eq:rcell}
\end{equation}
where $\langle \rho_p^2 \rangle = \int \rho_p^2(\boldsymbol{r})\mathrm{d}\boldsymbol{r}/V_{\rm{cell}}$ and $\langle \rho_p \rangle = \int \rho_p(\boldsymbol{r})\mathrm{d}\boldsymbol{r}/V_{\rm{cell}}$.
The WS cell volume $V_{\rm{cell}}$ depends on the spatial dimensionality $D$, defined as: 
\begin{equation}
V_{\rm{cell}}=\left\{
\begin{array}{ll}
\frac{4}{3}{\pi}R_{\text{W}}^{3} &\hspace{0.5cm} D=3 \ ,  \\
{\pi}aR_{\text{W}}^{2}           &\hspace{0.5cm} D=2 \ ,        \\
R_{\text{W}}{a^2}               &\hspace{0.5cm} D=1 \ .
\end{array}
\right.
\label{eq:vcell}
\end{equation}
where $D = 3$ corresponds to droplets and bubbles,
$D = 2$ to rods and tubes, and 
$D = 1$ to slabs.
Since rods, tubes, and slabs are infinite in extent for $D = 1$ and $2$, a finite length scale $a$ is introduced to ensure the volume is well-defined.

In the outer crust, only droplet structures are considered, as they are energetically favored in this region. Due to the large size of the WS cell at low densities and the computational constraints of numerical simulations, the WS cell is divided into two regions: a central core of radius $R_{\rm{in}}$, and a surrounding spherical shell ($R_{\rm{in}}<r<R_{\rm{W}}$).
In the shell region, electrons are assumed to have a constant density, and the total energy of the WS cell is given by:
\begin{equation}
E_{\mathrm{cell}}=\int_{\mathrm{core}}{\varepsilon}_{\mathrm{QMF,crust}}({\boldsymbol{r}})\mathrm{d}{\boldsymbol{r}}+\varepsilon_eV_{\rm{cell}}\ ,
\label{eq:TFe_outer}
\end{equation}
where $\varepsilon_e$ is the kinetic energy density of the uniformly distributed electrons in the shell. The transition between the outer and inner crust occurs at the density $\rho_{\rm{oi}}$, defined by the condition that the neutron chemical potential exceeds the neutron rest mass, indicating the onset of neutron drip from nuclei.

\subsection{Global properties of NSs 
}\label{sec:theory_MR}

The NS stable configuration in hydrostatic equilibrium can be obtained by solving the Tolman-Oppenheimer-Volkoff (TOV) equation~\cite{Tolman1939_PR055-364,Oppenheimer1939_PR055-374} for the pressure $P$ and the enclosed mass $m$,
\begin{eqnarray}
\frac{dP}{dr} &=& -\frac{[P(r) + \varepsilon(r)][m(r) + 4\pi r^3P(r)]}{r[r-2m(r)]} \ ,\nonumber\\
\frac{dm}{dr} &=& 4\pi r^2 \varepsilon(r) \ ,
\label{eq:TOV}
\end{eqnarray}
Given a central density $\rho_c$ at $r=0$, the TOV equation is
integrated outward until the pressure vanishes at $r=R$, which defines the radius $R$ of the NS. The total gravitational mass is then $M=m(R)$. 
To determine the core-crust boundary, we identify the core radius $R_{\mathrm{core}}$ by solving $P(R_{\mathrm{core}})=P_{\mathrm{cc}}$, where $P_{\mathrm{cc}}$ is the pressure at the crust-core transition, corresponding to the transition density $\rho_{\mathrm{cc}}$. The associated core mass is then given by $M_{\mathrm{core}}=m(R_{\mathrm{core}})$.

\subsection{Approximately unified EoSs based on \citet{2016arXiv161101357Z} 
}\label{sec:theory_Zdunik}

In this work, it is interesting to compare with the approximate method proposed by \citet{2016arXiv161101357Z} for determining the radius and mass of NSs, where the TOV equation is solved approximately in the crust by taking $m\approx M$ and $4\pi r^3P/m\ll1$. Through this method, a precise formula for the radius is obtained, which depends on only on $M_{\rm core}$ and $R_{\rm core}$, the baryon chemical potential at the crust-core interface $\mu_{\rm cc}$, and the baryon chemical potential on the crust surface $\mu_0$, i.e.,
\begin{equation}
    R=\frac{R_{\rm core}}{1-(\mu_{\rm cc}^2/\mu_0^2-1)(R_{\rm core}/M-1)} \ ,
\end{equation}
where the total mass 
\begin{equation}
    M=M_{\rm core}+M_{\rm crust}\ ,
\end{equation} 
with $M_{\rm crust}$ the crust mass and is given by
\begin{equation}
    M_{\rm crust}=\frac{4\pi P_{\rm cc}R_{\rm core}^4}{M_{\rm core}}(1-\frac{2M_{\rm core}}{R_{\rm core}})\ .
\end{equation}

\subsection{Cooling processes of isolated and transiently accreting NSs } \label{sec:theory_cooling}

While the static structure — such as masses, radii, and tidal deformabilities — has been extensively studied within the QMF model, the dynamic evolution of NSs remains a less explored but equally important aspect.
The thermal evolution of NSs is governed by the local energy balance and heat transport equations \cite{2006NuPhA.777..497P,1996NuPhA.605..531S},
\begin{align}
    & c_{v}\frac{\partial(Te^{\phi})}{\partial t}=-e^{2\phi}(q_{\nu}+q_{\mathrm{p}}-q_{\mathrm{heat}})-\frac{1}{4\pi r^2(1+z)}\frac{\partial(Le^{2\phi})}{\partial r}, \label{eq:energybalance} \\
    & Le^{2\phi} = -\frac{4\pi r^2\kappa e^{\phi}}{1+z}\frac{\partial(Te^{\phi})}{\partial r} \ . \label{eq:heattransport}
\end{align}
Here, $T$ and $L$ represent the internal temperature and luminosity of the star, respectively.
$\kappa$ and $c_{v}$ are the local thermal conductivity and specific heat, while $q_{\nu}$ and $q_{\mathrm{p}}$ are the local neutrino and photon emissivitiy, respectively. 
$q_{\mathrm{heat}}$ accounts for all possible energy sources contributing to the heating of the star.  
$\phi$ is the curvature profile of the star, and $1+z$ represents the surface gravitational redshift, with $1+z=(1-2M/R)^{-1/2}$.  

Current simulations of thermal evolution typically use a general relativistic framework, and several robust codes have been developed, such as \texttt{NSCool}~\footnote{http://www.astroscu.unam.mx/neutrones/
NSCool/} for the entire star, and \texttt{crustcool}~\footnote{https://github.com/andrewcumming/crustcool} for the crust.

For our simulations of the thermal evolution of isolated NSs, we employ the \texttt{NSCool} code, which incorporates various relevant cooling reactions: direct Urca (dUrca: $n\rightarrow p+l+{\bar\nu _{l}},~p+l\rightarrow n+\nu _{l}$) \cite{1991PhRvL..66.2701L}, modified Urca (mUrca: $n+N\rightarrow p+N+l+{\bar\nu _{l}},~p+N+l\rightarrow n+N+\nu _{l}$) \cite{1964PhRvL..12..413C,1979ApJ...232..541F}, nucleon-nucleon bremsstrahlung (NNB: $N+N\rightarrow N+N+\nu +\bar{\nu}$) \cite{1975PhRvD..12..315F}, Cooper pair breaking and formation processes (PBF: $N+N\rightarrow \left[ NN\right] +\nu +\bar{\nu},~[NN]\rightarrow N+N+\nu+\bar{\nu}$) \cite{1976ApJ...205..541F}. 
For simplicity, we neglect any possible heating in our cooling simulations of isolated NSs.
The NS EoS determines the stellar structure, as well as the effective masses and pairing gaps of baryons, and is therefore crucial for evaluating the specific heat and neutrino emissivity~\cite{2015SSRv..191..239P}. 
In our simulations, we use the neutron and proton $^1S_0$ critical temperatures from Refs. \cite{Wambach1993_NPA555-128} and \cite{Amundsen1985_NPA437-487}, respectively. 
The neutron $^3P_2$ critical temperature is based on the work of Ref. \cite{Page2004_ApJSupp155-623}, i.e. model maintains the same momentum dependence as model ``a'' but the specific values are rescaled to half of their original values (with maximum critical temperature $0.5\times10^9$ K).
To simplify, we treat non-spherical nuclear clusters as spherical. With our unified EoS and NS structure as input, we solve the thermal evolution equations (\ref{eq:energybalance}) and (\ref{eq:heattransport}) using initial profiles for $T(r,t=0)$ and $L(r,T=0)$, along with boundary conditions for $T(r,t)$ and $L(r,t)$ at $r=0$ and $r=r_{\mathrm{b}}$. 
Here, $r_{\mathrm{b}}$ represents the radius of the bottom of the envelope, and the relationship between the temperature $T_{\mathrm{b}}$ at $r=r_{\mathrm{b}}$ and the surface effective temperature $T_{\mathrm{e}}$ follows the so-called $T_{\mathrm{e}}(T_{\mathrm{b}})$ relation, as described in Ref. \cite{Page2004_ApJSupp155-623}.

For an X-ray transient, the effective temperature typically remains constant due to the balance between deep crustal heating and core neutrino emission~\cite{1990ApJ...362..572M,2001ApJ...548L.175C}. During accretion, the thermal nuclear burning of light elements in the accretion layer heats the crust out of thermal equilibrium with the core, leading to an X-ray burst. 
After accretion, the crust returns to thermal equilibrium with the core via crustal cooling~\cite{1998ApJ...504L..95B,2002ApJ...580..413R}. The crustal cooling depends on the composition, the EoS, and neutron $^1S_0$ pairing gap in the crust, all of which are determined through self-consistent calculations in this work. 
The employed heat-blanketing envelope consists of pure He down to $10^9$ g cm$^{-2}$ and pure Fe down to the bottom of the envelope ($10^{12}$ g cm$^{-2}$)~\cite{2009ApJ...698.1020B,2024EPJA...60..116A}. 
We set the local mass accretion rate as $\Dot{m}=0.1\Dot{m}_{\mathrm{Edd}}$with the local Eddington mass accretion rate $\Dot{m}_{\mathrm{Edd}}=8.8\times10^4$ g cm$^{-2}$ s$^{-1}$ \cite{2017ApJ...839...95D}. For our simulations, we neglect all possible neutrino emissions from the crust.
The remaining stellar properties to be determined include the NS mass, the core temperature $T_{\mathrm{core}}$, the temperature $T_{\mathrm{t}}$ at the bottom of the envelope during accretion, and the impurity parameter $Q_{\mathrm{imp}}$ of the crust. 

To determine the optimal stellar properties, we carry out a case study using the \texttt{crustcool} code. We run Markov chain Monte Carlo simulations for the source KS 1731-260 \cite{Merritt2016_ApJ833-186}, using typical values for the NS mass and EoSs calculated by the QMF and RMF models, with symmetry energy slope values $L_0$ ranging from 40 to 80 MeV.

\subsection{Viscosity and r-mode instability}\label{sec:theory_viscosity}

Bulk and shear viscosities are usually considered as the primary dissipation mechanisms for r-modes and other oscillation modes in NSs. 
Bulk viscosity arises when r-mode oscillations drive the dense matter out of $\beta$-equilibrium. 
As the system attempts to restore chemical equilibrium through weak interactions, energy is dissipated. 
At high temperatures ($>10^9\ \mathrm{K}$), bulk viscosity from the modified Urca (mUrca) or direct Urca (dUrca) processes dominates the dissipation. 
At lower temperatures ($<10^9\ \mathrm{K}$), shear viscosity due to neutron and electron scattering becomes the dominant damping mechanism.

The total bulk viscosity can be simply written as a sum of the partial bulk viscosities associated with various (modified and direct) Urca processes~\citep {2000A&A...357.1157H,2001A&A...372..130H,2012PhRvC..85d5808V}:,
\begin{eqnarray}
\xi &=& \xi_{\text{mUrca}} + \xi_{\text{dUrca}}\nonumber\\
    &=& \sum_{Nl}\frac{|\lambda_{Nl}|}{\omega^2} \Big| \frac{\partial P}{\partial X_l} \Big| \frac{\partial \eta_l}{\partial \rho_B} 
    + \sum_{l}\frac{|\lambda_{l}|}{\omega^2} \Big| \frac{\partial P}{\partial X_l} \Big|\frac{\partial \eta_l}{\partial \rho_B} \ .
\label{eq:Urca}
\end{eqnarray}
Here, $\omega$ is the frequency of the oscillation mode,
$X_l = \rho_l/\rho_B$ represents the lepton (electron or muon) fraction, ~$\eta_l = \mu_n - \mu_p - \mu_l$, with $\mu_i$ being the chemical potential of particle species~$i$, and~$\lambda_{Nl}$ and~$\lambda_{l}$ determine the difference of
the rates of the direct and inverse reactions of a given Urca process:~$\Gamma_{Nl} - \bar{\Gamma}_{Nl} = -\lambda_{Nl}\eta_l$ for the mUrca processes, and~$\Gamma_l - \bar{\Gamma}_l =  -\lambda_{l}\eta_l$ for the dUrca processes~\citep{2000A&A...357.1157H,2001A&A...372..130H,2012PhRvC..85d5808V}.

The partial bulk viscosity of $npe\mu$ matter arising from a nonequilibrium dUrca process is given by~\citep{2000A&A...357.1157H}:
\begin{eqnarray}   
    \xi_{l}^{\text{dUrca}}&=&8.553\times10^{24}\frac{M_{n}^{*}}{M_n}\frac{M_p^{*}}{M_p}\Big(\frac{\rho_e}{\rho_0} \Big)^{1/3}\nonumber\\
    &\times&\frac{T_9^4}{\omega_4^2}\Big(\frac{C_l}{100\ \mathrm{MeV}} \Big)^2\Theta_{npl}\ \mathrm{g\ cm^{-1}\ s^{-1}}\ ,
    \label{eq:DBulk}
\end{eqnarray}
where $\omega_4=\omega/(10^4~\mathrm{s^{-1}})$, $T_9=T/(10^9~\mathrm{K})$. 
The angular frequency $\omega$ of the r-mode is given by $\omega=-m\Omega+2m\Omega/l(l+1)$, with $\Omega$ being the stellar angular velocity~\citep{2003CQGra..20R.105A}. 
In this work, we focus only on r-modes with angular quantum number $l=2$ and azimuthal projection $m=2$. 
The coefficient $C_l$ is defined as:
\begin{eqnarray}
    C_l=4(1-2Y_p)\rho_B\frac{\mathrm{d}S(\rho_B)}{\mathrm{d}\rho_B}-\frac{c^2k_{F_l}^2}{3\mu_l}\ ,
\end{eqnarray}
where $S(\rho_B)$ is the symmetry energy at $\rho_B$. The step function $\Theta_{npl}$ equals 1 if the dUrca process opens for $k_{F_n}<(k_{F_l}+k_{F_p})$ and equals 0 otherwise. 
For mUrca processes, the bulk viscosities from different channels are given by~\citep{2001A&A...372..130H}:
\begin{align}
    \xi_{ne}^{\text{mUrca}}&=1.49\times10^{19}\Big(\frac{M_n^{*}}{M_n} \Big)^{3}\frac{M_p^{*}}{M_p}\Big(\frac{\rho_p}{\rho_0} \Big)^{1/3}\nonumber\\
    &\times\Big(\frac{C_e}{100\mathrm{MeV}} \Big)^2\frac{T_9^6}{\omega_4^2}\alpha_n\beta_n\ \mathrm{g\ cm^{-1}s^{-1}} \,
    \\
    \xi_{pe}^{\text{mUrca}}&=\xi_{ne}^{\text{mUrca}}\Big(\frac{M_p^{*}}{M_n^{*}} \Big)^2\frac{(3k_{F_p}+k_{F_e}-k_{F_n})^2}{8k_{F_p}k_{F_e}}\Theta_{pe} \,
    \\
    \xi_{n\mu}^{\text{mUrca}}&=\xi_{ne}^{\text{mUrca}}\Big(\frac{k_{F_\mu}}{k_{F_e}} \Big)\Big(\frac{C_\mu}{C_e} \Big)^2 \,
    \\
    \xi_{p\mu}^{\text{mUrca}}&=\xi_{ne}^{\text{mUrca}}\Big(\frac{C_\mu M_p^{*}}{C_e M_n^{*}} \Big)^2\frac{(3k_{F_p}+k_{F_\mu}-k_{F_n})^2}{8k_{F_p}k_{F_\mu}}\\
    &\times\Big(\frac{k_{F_\mu}}{k_{F_e}} \Big)\Theta_{p\mu} \ ,
\end{align}
where the step function $\Theta_{pl}=1$ if $k_{F_n}<(k_{F_l}+3k_{F_p})$ and $0$ otherwise.

Shear viscosity $\eta$ plays a key role in damping r-mode oscillations in cooler NSs. 
It arises from momentum transport due to particle scattering. 
In general, the total shear viscosity can be approximated as a sum of partial contributions from neutron scattering and electron scattering. 
The electron shear viscosity $\eta_e$ results primarily from electron interactions with other charged particles via electromagnetic forces, whereas the neutron contribution $\eta_n$ is governed by neutron-neutron and neutron-proton collisions mediated by the strong interaction. 
In this work, we adopt the following parameterizations for the neutron and electron shear viscosities~\citep{2008PhRvD..78f3006S,1987ApJ...314..234C},
\begin{equation}
\eta_n =2\times10^{18}(\rho_{15})^{9/4}T_9^{-2}~\rm{g~cm^{-1}~s^{-1}}, 
\label{eq:shearn}
\end{equation}
\begin{equation}
\eta_e =8.33\times10^{19}(Y_p\rho_{15})^{14/9}T_9^{-5/3}~\rm{g~cm^{-1}~s^{-1}}. 
\label{eq:sheare}
\end{equation}
where $\rho_{15}=\rho/(10^{15}\,{\rm g\,cm^{-3}})$. 
Both contributions, $\eta_n$ and $\eta_e$, are taken into account in our analysis. 

The time dependence of an r-mode oscillation is described by $e^{i\omega t-t/\tau}$, where $\tau$ is an overall timescale incorporating both the exponential growth driven by the Chandrasekhar-Friedman-Schutz mechanism and the decay due to viscous damping. 
The timescale is given by~\citep{1998PhRvL..80.4843L}
\begin{eqnarray}
\frac{1}{\tau(\nu,T)} = -\frac{1}{\tau_{\text{GW}}(\nu,T)} + \frac{1}{\tau_{\xi}(\nu,T)} + \frac{1}{\tau_{\eta}(\nu,T)}. 
\end{eqnarray}
where $\tau_{\text{GW}}$ is the time scale of the growth of an r-mode due to the emission of gravitational waves~\cite{1998PhRvL..80.4843L}:
\begin{eqnarray}
\frac{1}{\tau_{\text{GW}}} &=& \frac{32\pi G\Omega^{2l+2}}{c^{2l+3}}\frac{(l-1)^{2l}}{[(2l+1)!!]^2}\Big(\frac{l+2}{l+1}\Big)^{2l+1}\nonumber\\
&\times& \int_{0}^{R}\rho r^{2l+2} dr,
\end{eqnarray}
and the dissipative timescales due to bulk and shear viscosity are: 
\begin{eqnarray}
\frac{1}{\tau_{\xi}} &=& \frac{4\pi}{690}\Big(\frac{\Omega^2}{\pi G \bar{\rho}}\Big)R^{2l-2}\Big[\int_0^R \rho r^{2l+2}dr\Big]^{-1}\nonumber\\
&\times& \int_{0}^{R}\xi \Big(\frac{r}{R}\Big)^6 \Big[1+0.86\Big(\frac{r}{R}\Big)^2\Big]r^2 dr,
\end{eqnarray}
\begin{eqnarray}
\frac{1}{\tau_{\eta}} &=& (l-1)(2l+1)\Big[\int_0^R \rho r^{2l+2}dr\Big]^{-1}\nonumber\\
&\times& \int_{0}^{R}\eta r^{2l}dr,
\end{eqnarray}
where $\bar{\rho}\equiv M/(4\pi R^3/3)$ is the average density of a nonrotating star, and $\Omega=2\pi\nu$ is the angular velocity. 
If $1/\tau<0$, the r-mode becomes unstable and grows exponentially. Conversely, if $1/\tau>0$, the mode is damped. 
For a given stellar temperature $T$, the critical frequency $\nu_c$ is defined as the smallest root of $1/\tau(\nu_c, T)=0$. 
Similarly, for a given frequency $\nu$, the critical temperature $T_c$ satisfies $1/\tau(\nu, T_c)=0$. 
Thus, according to equation $1/\tau(\nu, T)=0$, we can delineate the r-mode instability region in the $\nu$-$T$ plane.
Therefore, in determining the critical temperature of r-mode instability for cool stars, it is reasonable to consider shear viscosity as the leading dissipation channel.
This is especially relevant since all observed low-mass X-ray binaries (LMXBs) exhibit temperatures around $10^8~\mathrm{K}$, and old NSs are typically even cooler~\citep{2011PhRvL.107j1101H,2021ApJ...910...62Z}.  

\section{Results and discussions}\label{sec:result}
In this section, we will discuss and compare the performances of QMF and RMF models in describing NS matter and the global and dynamical properties of the NSs. 

\subsection{NS EoS, composition, and global properties: QMF vs. RMF}\label{sec:result_EOS}

The microscopic structure of crustal matter significantly influences thermal properties such as specific heat and thermal conductivity, thereby playing a crucial role in the thermal evolution of NSs (as detailed below in Sec.~\ref{sec:result_cooling}). 
Table \ref{tab:transdensity} presents the transition densities between the inner and outer crust ($\rho_{\text{oi}}$), as well as between the crust and core ($\rho_{\text{cc}}$), calculated using different effective interactions within two models. The dependence of these transition densities on the slope parameter $L_0$ in the QMF model is consistent with previous studies: A larger $L_0$ leads to a higher $\rho_{\text{oi}}$ but a lower $\rho_{\text{cc}}$. Compared to the RMF model, the QMF model predicts slightly lower transition densities, within computational uncertainty, although $\rho_{\text{cc}}$ are nearly identical at smaller $L_0$.

Our analysis focuses on the evolution of the total nucleon number $A_{\rm{cell}}$, the nucleon number of the ion $A_{\rm{ion}}$, the proton number $Z$, and the density of the free neutron gas $\rho_n^{\infty}$ within the WS cells in the crust. 
In the outer crust, these quantities calculated by the QMF model are slightly larger than those from the RMF model. For instance, the outermost nucleus predicted by the QMF model with $L_0=40$ MeV is $^{53}$Cr, whereas the RMF model yields $^{48}$Ti.
At the outer-inner crust interface, the innermost nuclei are $^{103}$Ge and $^{91}$Cu for the QMF and RMF models, respectively. 
These isotopes are obtained by rounding the nucleon numbers at the crust boundaries. 
The stronger $\sigma$-meson field in the QMF model at sub-saturation densities alters nuclear stability through the derivative term in Eq.~(\ref{eq:motion4}), leading to heavier stable nuclei. 
This suggests that the neutron dripline is reached at a lower density in the QMF model, corresponding to a lower $\rho_{\rm{oi}}$. 
Since variations in $L_0$ affect only one proton or a few nucleons, its influence on the outer crust structure is relatively minor.

\begin{figure}
{\centering
{\includegraphics[width=1.0\linewidth]{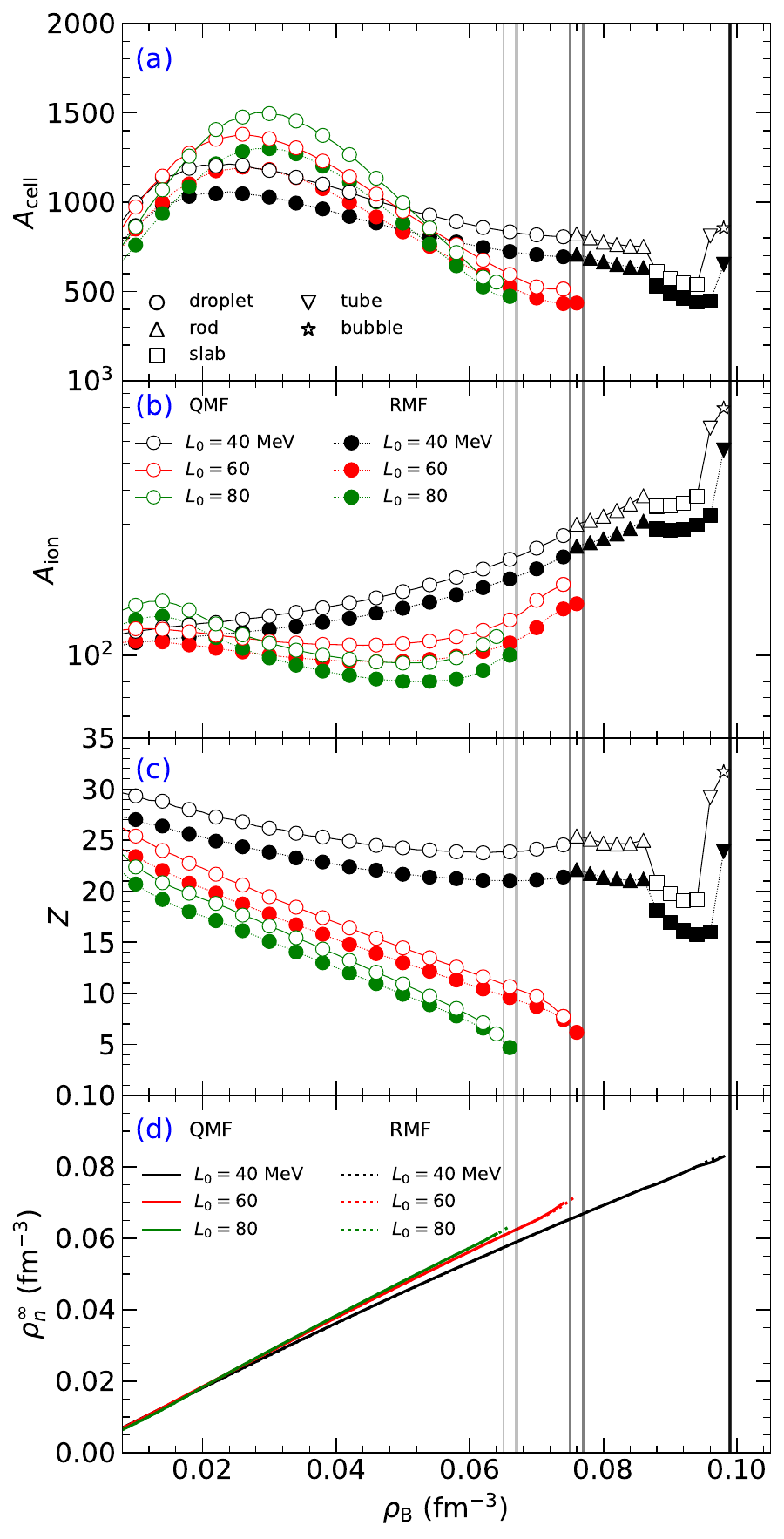}}}
\caption{\label{fig:crust_composition}(a) Total nucleon number $A_{\text{cell}}$, (b) nucleon number of the ion $A_{\text{ion}}$, (c) proton number $Z$, and (d) neutron gas density $\rho_n^{\infty}$ as functions of the baryon density $\rho_B$ in the inner crust. 
Open and closed circles represent results from the QMF and RMF models, respectively. 
Black, red, and green lines correspond to calculations with $L_0=40,~60,~80$ MeV, respectively. 
Five types of pasta phases are considered and are indicated using different markers in the figure.
Thin and thick vertical lines denote the crust–core transition densities predicted by the QMF and RMF models, respectively.
The gray scale of the vertical lines gradually deepens as $L_0$ decreases, e.g., the darkest and lightest vertical lines corresponding to the crust-core transition densities calculated using effective interactions with $L_0=40$ and 80 MeV, respectively.
}
\end{figure}

\begin{figure}
{\centering
{\includegraphics[width=1.0\linewidth]{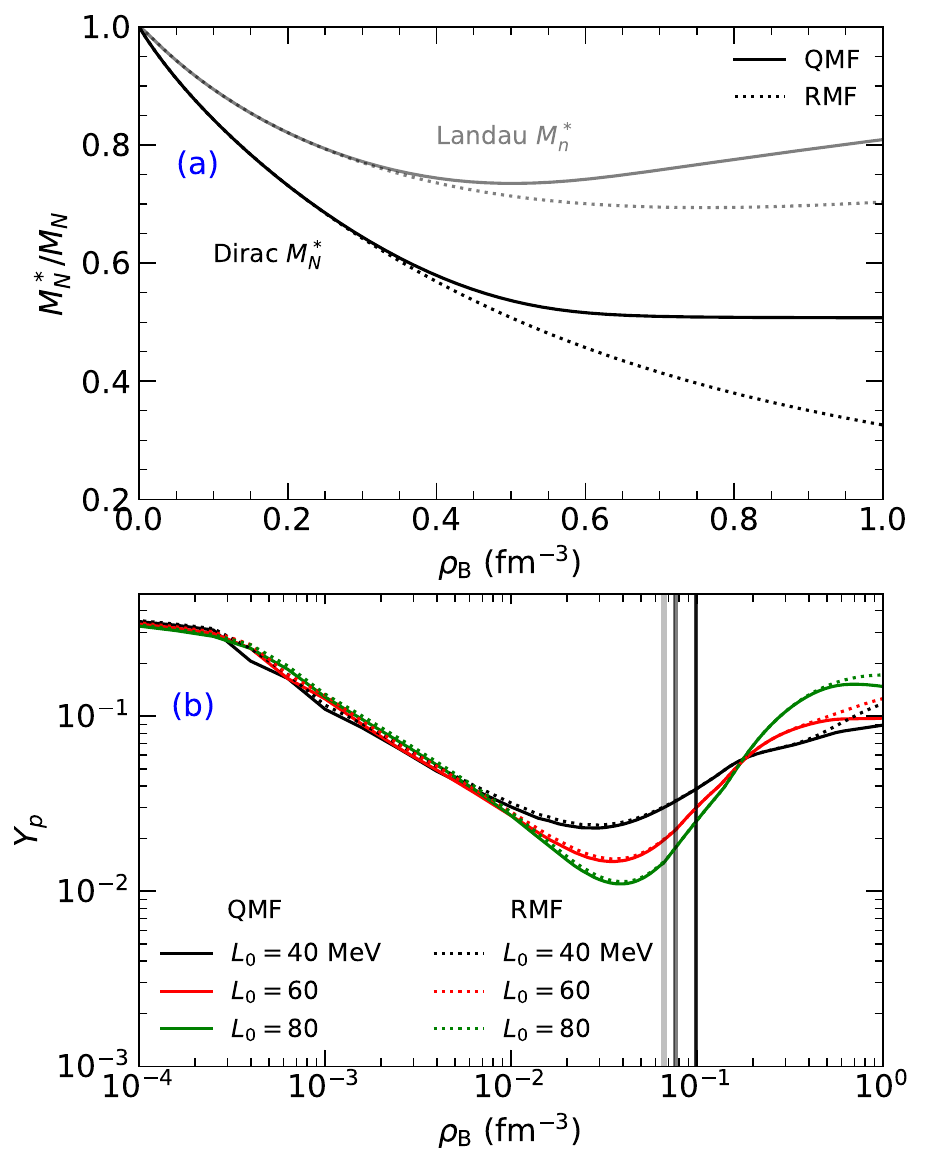}}}
\caption{\label{fig:core_composition}(a) Effective nucleon mass in uniform nuclear matter calculated using the QMF and RMF models. The Dirac effective mass depends only on the isoscalar field, while variations in $L_0$ are associated with the isovector interaction in both models; therefore, $L_0$ has no impact on the Dirac effective mass. (b) Proton fraction in the inner crust and core for the QMF and RMF models, shown for three values of the symmetry energy slope: $L_0=40,~60,~80$ MeV.
}
\end{figure}

As shown in Fig. \ref{fig:crust_composition}(a)--(c), the inner crust structure exhibits a stronger dependence on both $L_0$ and the model used.
For all structural indicators ($A_{\rm{cell}}$, $ A_{\rm{ion}}$, and $Z$), the QMF model consistently yields higher values than the RMF model—again reflecting the same mechanism observed in the outer crust. 
This implies that, at a given density, both the WS cell and nuclear cluster are larger in the QMF model, which could significantly affect thermal conductivity, particularly when finite-size effects are considered. 
The dependence of the microscopic structure on $L_0$ is more complex but exhibits similar trends in both models. 
For example, at $L_0=40$ MeV, non-spherical pasta structures appear. The QMF model produces all considered pasta structures, , including the bubble phase, which is not realized in the RMF model.
At higher $L_0$ values (60 and 80 MeV), only droplet structure is obtained. 
Fig. \ref{fig:crust_composition}(d) shows the evolution of $\rho_n^{\infty}$ as a function of $\rho_B$ in the inner crust.
At a given $\rho_B$, $\rho_n^{\infty}$ increases with $L_0$.
The values of $\rho_n^{\infty}$ predicted by the QMF and RMF models are nearly identical, indicating that the larger nucleon numbers in the QMF model are compensated by correspondingly larger WS cell sizes.
Furthermore, since the effective masses predicted by both models are nearly the same at low densities (see also Fig.~\ref{fig:core_composition}(a)), the neutron pairing properties in the inner crust show no model dependence under the standard BCS approximation. Thus, it is justified to adopt the same neutron pairing gap for both the QMF and RMF models in the crust.

\begin{table}
  \centering
  \caption{Calculated outer-inner crust transition density $\rho_{\rm{oi}}$ and crust-core transition density $\rho_{\rm{cc}}$ for QMF and RMF models with different $L_0$. The corresponding numerical uncertainties are $10^{-4}$ fm$^{-3}$ for $\rho_{\rm{oi}}$ and 0.002 fm$^{-3}$ for $\rho_{\rm{cc}}$.
  }\label{tab:transdensity}
  \setlength\tabcolsep{5pt}
  \begin{tabular}{lcccccc}
    \hline
    \hline
     & \multicolumn{3}{c}{QMF} & \multicolumn{3}{c}{RMF}    \\
     \cmidrule(r){2-4} \cmidrule(r){5-7} 
     $L_0$ (MeV) & 40 & 60 & 80 & 40 & 60 & 80 \\ 
    \hline
    $\rho_{\rm{oi}}$ ($10^{-4}$ fm$^{-3}$) & 2.517 & 2.663 & 3.525 & 2.649 & 3.314 & 3.671 \\
    $\rho_{\rm{cc}}$ (fm$^{-3}$) & 0.099 & 0.075 & 0.065 & 0.099 & 0.077 & 0.067 \\
    \hline     \hline
  \end{tabular}
\end{table}

Regarding the composition of the NS core, our primary focus is on the nucleon effective mass $M_N^*$ and the proton fraction $Y_p$, as these quantities are closely tied to several important astrophysical processes. For example, the dUrca process in the $np$ channel requires a proton fraction exceeding approximately 1/8, and its specific neutrino emissivity depends on the effective mass~\cite{1991PhRvL..66.2701L}. Additionally, the scattering of electrons off vortex cores magnetized by entrainment leads to core energy dissipation, where the drag parameter is determined by the proton effective mass and proton fraction \cite{2006MNRAS.368..162A}. As demonstrated in our previous work \cite{2023PhRvC.108b5809Z} and in Fig. \ref{fig:core_composition} (a), the main difference between the RMF and QMF models lies in the behavior of the effective mass at high densities. In the QMF model, the effective mass tends to approach a constant value (approximately 0.5$M_N$ in the present study), whereas in the RMF model, it decreases monotonically with increasing density. At lower densities, the two models yield similar results.  In the QMF framework, the effective mass is not solely determined by the background scalar field but also includes contributions from the chiral condensate and pion cloud. For further details, see \cite{2018ApJ...862...98Z}.

This model-dependent behavior of the effective mass directly leads to discrepancies in the proton fraction at high densities. As illustrated in Fig. \ref{fig:core_composition}(b),  the QMF model predicts a significantly suppressed proton fraction in the high-density regime.
The onset of this suppression coincides with the density range (around 0.3--0.4 fm$^{-3}$) where the effective mass predicted by the two models begins to diverge. 
The reduction in proton fraction at high effective mass arises from the constraints of $\beta$-equilibrium and baryon number conservation conditions.
Regarding the dependence on the symmetry energy slope parameter $L_0$, a larger $L_0$ generally favors more symmetric nuclear matter, and thus leads to an increase in the proton fraction above saturation density.
However, this $L_0$-dependence does not qualitatively alter the differences between the QMF and RMF models.

\begin{figure}
{\centering
{\includegraphics[width=1.0\linewidth]{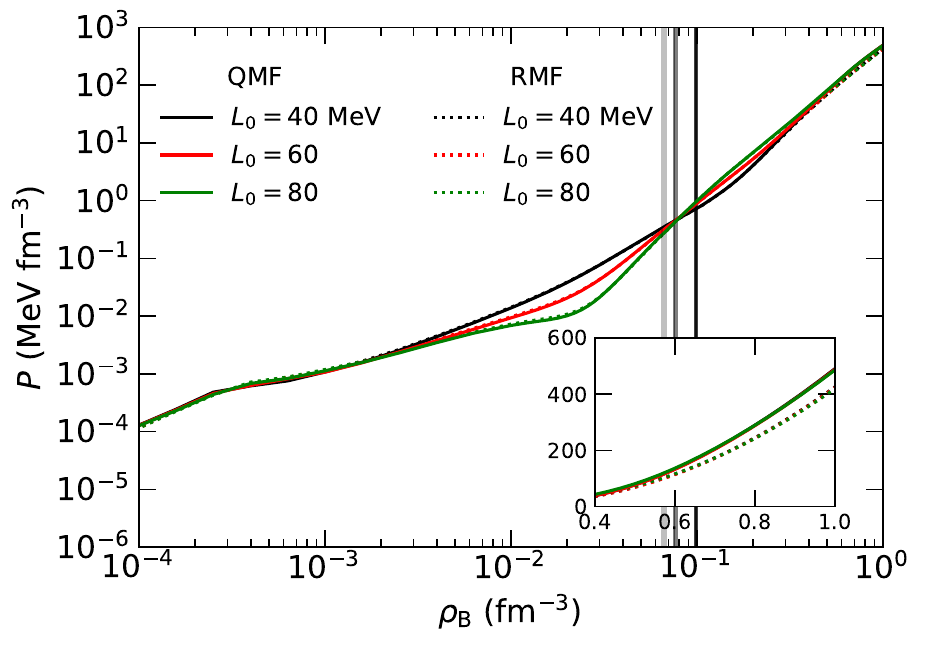}}}
\caption{\label{fig:EoS} Unified EoSs calculated using the QMF and RMF models for $L_0=40,~60,~80$ MeV. The inset in the lower right corner magnifies the EoS details at high densities ($>0.4~\rm{fm}^{-3}$).
}
\end{figure}

By integrating the energy density functional obtained from the QMF model with different $L_0$ values over the internal structure of NSs, we obtain the EoS for NS matter. 
Fig. \ref{fig:EoS} shows the pressure as a function of baryon number density. The QMF model predicts higher pressures in the high-density regime, as seen in the inset of Fig.~\ref{fig:EoS},  while the influence of $L_0$ is more pronounced in the intermediate-density region. 
Specifically, a larger $L_0$ results in a higher pressure above the crust-core transition density. The pressure predicted by the QMF model at high densities is greater than that predicted by the RMF model, which can be attributed to the larger kinetic term of pressure induced by the higher effective mass at high densities.

\begin{figure*}
\vspace{-0.5cm}
{\centering
{\includegraphics[width=0.75\linewidth]{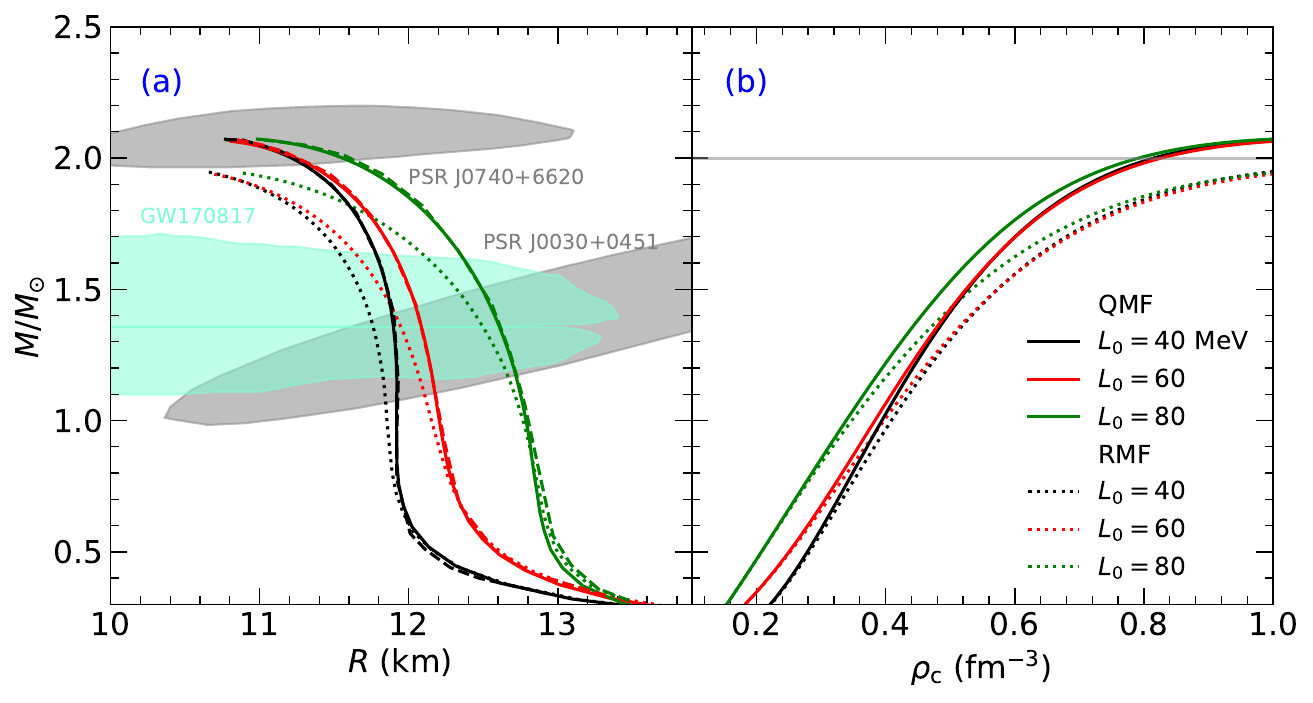}}}
\caption{(a) Mass–radius relations and (b) density profiles calculated using the QMF and RMF models for $L_0=40,~60,~80$ MeV. The dashed lines represent results based on the approximately unified EoS from \citet{2016arXiv161101357Z} for the QMF model. Shaded regions indicate observational constraints on the mass–radius relation: from the tidal deformability measurement of GW170817 \citep{2017PhRvL.119p1101A}, and from NICER observations of PSR J0030+0451 \citep{2024ApJ...961...62V} and PSR J0740+6620 \citep{2024ApJ...974..294S}. All these measurements are presented at the 90\%  confidence level. 
}\label{fig:MR}
\end{figure*}

\begin{figure}
{\centering
{\includegraphics[width=1.0\linewidth]{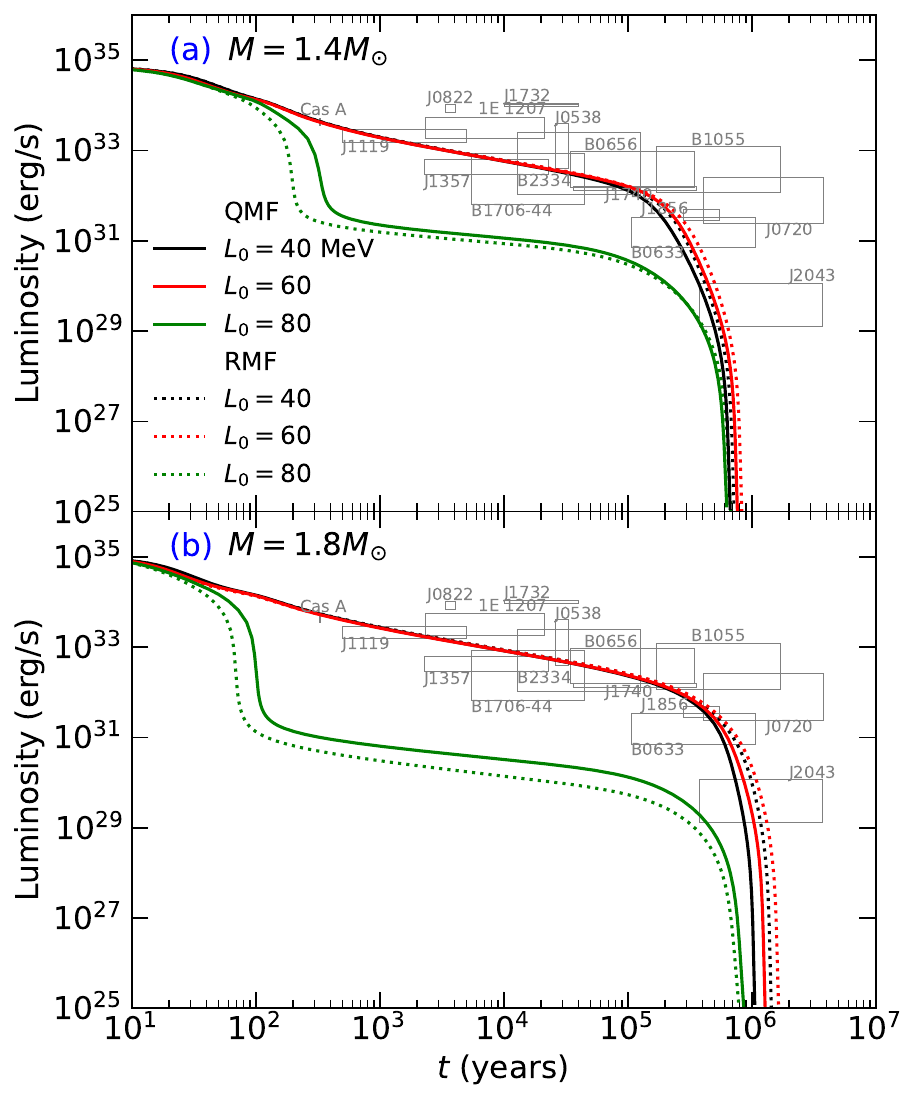}}}
\caption{Cooling curves of the isolated NS based on the QMF and RMF models with different $L_0$ values for NS masses (upper panel) 1.4$M_{\odot}$ and (lower panel) 1.8$M_{\odot}$. The observation data are taken from Table I of Ref. \cite{Beloin2018_PRC97-015804}. 
} \label{fig:Lumi}
\end{figure}

Using the unified EoSs derived in this work as input, the $M$-$R$ relations and density profiles of NSs are obtained by solving the TOV equations. The results are shown in Fig. \ref{fig:MR}. 
In Fig. \ref{fig:MR}(a),
the higher pressure predicted by the QMF model at high densities leads to a larger maximum mass compared to the RMF model, with a difference of approximately $\Delta M \approx 0.13 M_{\odot}$. 
The symmetry energy slope parameter $L_0$ governs the behavior of the EoS in the intermediate-density regime and consequently affects t
he radii of typical NSs with $M=1.4M_{\odot}$. In our present study, regardless of the adopted model, a larger $L_0$ results in larger stellar radii.
Fig.~\ref{fig:MR}(b) shows that the dependence of NS mass on $L_0$ is more pronounced at lower central densities and becomes less prominent at higher densities. In contrast, the dependence on the underlying model becomes more pronounced at higher central densities in our calculations.

The global properties calculated using the approximately unified EoS for the QMF model, following the method of \citet{2016arXiv161101357Z}, are also shown in Fig.~\ref{fig:MR}(a) (dashed lines).
These results are in excellent agreement with our fully unified EoS calculations for NS masses above 0.5 $M_{\odot}$ across different values of $L_0$. 
A similar level of agreement is observed within the RMF model as well, indicating that the approximate method provides a highly accurate representation of the exact unified EoS.

In the following sections, we will investigate how the model- and $L_0$-dependence of the EoS and internal composition affect the dynamical behavior of NSs. 

\subsection{NS cooling from the insight of interior}\label{sec:result_cooling}

The mass and radius of an NS, together with its EoS and internal composition, fundamentally determine its thermal evolution. The cooling of isolated NSs is governed by neutrino emission from the core and photon emission from the surface, with neutrino processes dominating the thermal evolution for the first $10^5$ years. 
This early cooling phase is particularly important for probing the internal thermal structure of NSs. 
In Fig. \ref{fig:Lumi}, we show the cooling curves predicted by the QMF and RMF models for isolated NSs with masses of 1.4$M_{\odot}$ and 1.8$M_{\odot}$, along with observational data on surface luminosities from several sources for comparison.
For $L_0=40$ and $60$ MeV, the cooling curves are largely insensitive to both the model and $L_0$, and show good agreement with observational data. However, for $L_0=80$ MeV, both models predict the onset of the dUrca process in the core, which leads to significantly faster cooling. In this case, the QMF model predicts a longer thermal relaxation time than the RMF model, meaning the most rapid drop in surface luminosity occurs at a later time.

To further understand the cooling behavior, we next examine how the thermal structure and thermal properties of NSs depend on the model.

\begin{figure}
\centering
\includegraphics[width=0.45\textwidth]{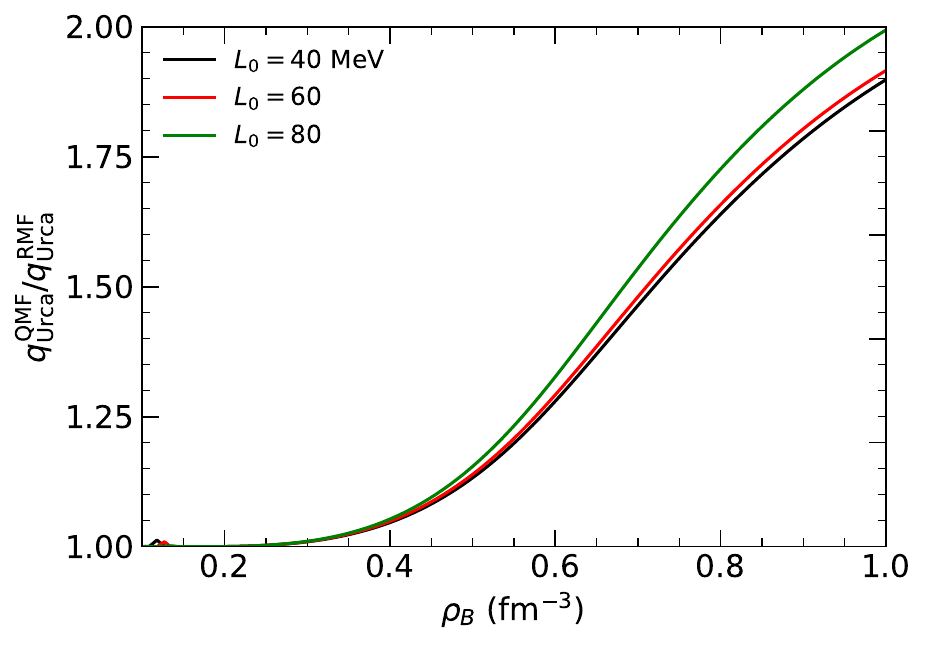}
\caption{Ratio of neutrino emissivities, summed over the mUrca and bremsstrahlung processes, calculated using the QMF model to those calculated using the RMF model.
}
\label{fig:Urca_emissivity}
\end{figure}

In the absence of the dUrca process, the dominant neutrino emission mechanisms are the mUrca, bremsstrahlung, and PBF processes. In our simulations, neutron $^3P_2$ superfluidity plays an important role in core cooling and we adopt a fixed maximum critical temperature of $0.5 \times 10^9$ K following Ref.~\cite{2011PhRvL.106h1101P}. 
Before the core temperature drops below the critical temperature, the mUrca and bremsstrahlung dominate neutrino emission, while the PBF process takes over afterward.
Since the nucleon effective mass and the Fermi momentum of nucleons and leptons influence the neutrino emissivity of the mUrca and bremsstrahlung, the larger effective mass and enhanced isospin asymmetry predicted by QMF at high densities modify the neutrino emissivity. In the cooling simulations, we need to convert the (Dirac) effective mass to the Landau effective mass via $M_{\mathrm{L},N}^* = \sqrt{k_{\rm{F}}^2 + M_{N}^{*2}}$. 
We calculate the total neutrino emissivity from mUrca and bremsstrahlung under both the QMF and RMF models and present the ratio $q_{\rm{Urca}}^{\rm{QMF}} / q_{\rm{Urca}}^{\rm{RMF}}$ as a function of baryon density in Fig.~\ref{fig:Urca_emissivity}. 
The results indicate that the QMF model enhances the total emissivity, with the enhancement increasing with density. However, even at very high densities ($\rho_B\approx1$ fm$^{-3}$), the ratio remains below 2, i.e., $q_{\rm{Urca}}^{\rm{QMF}} / q_{\rm{Urca}}^{\rm{RMF}} < 2$. 
Assuming that the pairing gap remains unchanged despite the altered effective mass, the ratio of neutron $^3P_2$ PBF neutrino emissivity between the QMF and RMF models is approximately given by $q_{\rm{PBF}}^{\rm{QMF}} / q_{\rm{PBF}}^{\rm{RMF}} \approx M_{\mathrm{L},n}^{*\rm{QMF}} / M_{\mathrm{L},n}^{*\rm{RMF}}$. 
From the lower panel of Fig. \ref{fig:core_composition}, we obtain $q_{\rm{PBF}}^{\rm{QMF}} / q_{\rm{PBF}}^{\rm{RMF}} < 1.15$.
Therefore, the QMF model increases the neutrino emissivity by less than a factor of 2, which does not lead to an order-of-magnitude change in the neutrino emissivity. As a result, in the absence of the dUrca process, the cooling curves predicted by the QMF and RMF models are nearly indistinguishable.

When $L_0$ is sufficiently small to prevent the occurrence of the dUrca processes in the core, its impact on $L_0$ on the cooling curves is minimal, as the effective mass remains largely insensitive to $L_0$. However, as $L_0$ increases, the associated rise in proton fraction can trigger the onset of the dUrca process, as evident in the cooling curves for $L_0=80$ MeV shown in Fig. \ref{fig:Lumi}. 
Since the neutrino emissivity of the dUrca process is approximately $10^5$ times higher than that of the mUrca process, this leads to rapid cooling, resulting in significant deviations from observational data.
The thermal relaxation time of an NS is marked by a sharp drop in surface temperature or luminosity.We find that the QMF model predicts a longer thermal relaxation time than the RMF model. Previous studies ~\cite{Sales2020_A&A642-A42,2025ApJ...987....6T} have shown that NSs with a small dUrca-allowed region (i.e., a small dUrca core) can exhibit delayed thermal relaxation, with smaller dUrca cores corresponding to longer relaxation times.
To quantify this, we compute
the fractional volume of the dUrca core relative to the total core volume, denoted $f_V^{\rm{dUrca}}$.
For an NS with mass $1.4 M_{\odot}$, the QMF and RMF models yield $f_V^{\rm{dUrca}} = 2.6\%$ and $f_V^{\rm{dUrca}} = 7.8\%$, respectively.
For a more massive star with $1.8 M_{\odot}$, the corresponding values increase to $f_V^{\rm{dUrca}} = 24.7\%$ and $f_V^{\rm{dUrca}} = 33.7\%$. 
These results clearly indicate that the QMF model predicts a smaller dUrca core.
This difference arises because the larger effective mass in the QMF model suppresses the proton fraction, thereby shifting the onset density for the dUrca process to higher values compared to the RMF model. As a result, for NSs of the same mass, the QMF model yields a smaller dUrca core and a longer thermal relaxation time. In short,  the QMF and RMF models show distinguishable behavior primarily in the rapid cooling driven by the dUrca process.

In our cooling simulations, we used self-consistently equilibrium compositions and same critical temperatures in both the QMF and RMF models. As we mentioned earlier, it is reasonable to adopt the same neutron $^1S_0$ critical temperature in the crust. Nevertheless, the proton $^1S_0$ and neutron $^3P_2$ critical temperatures in the core may differ between the two models due to variations in the proton fraction and effective mass. 
Taking proton $^1S_0$ superfluidity as an example, the larger the Fermi momentum and effective mass, the higher the critical temperature. As shown in Fig.~\ref{fig:core_composition}, compared to the RMF model, the QMF model predicts a lower Fermi momentum but a higher effective mass for protons in the core.  These opposing effects suggest that the difference in the predicted proton critical temperatures between the two models is likely to be limited. The calculation of the neutron $^3P_2$ critical temperature is more complex. This study primarily focuses on the differences between the unified EoSs constructed by the two models and their manifestations in the static and dynamic properties of NSs. A fully self-consistent determination of nucleon critical temperatures in the core lies beyond the scope of the present work but will be pursued in future studies.

\begin{table*}[htbp]
\centering
\renewcommand{\arraystretch}{1.2}
\caption{Optimal stellar properties for the cooling of transiently accreting NSs  under the QMF and RMF models with varying values of $L_0$. 
The listed properties include the corresponding NS radius, as well as the fitted core temperature
$T_{\mathrm{core}}$, the temperature at the bottom of the envelope during accretion, $T_{\mathrm{t}}$ at the bottom of the envelope during accretion, and the crustal impurity parameter $Q_{\mathrm{imp}}$.
}
\label{tab:shorttermcooling}
\setlength\tabcolsep{10.0pt}
\begin{tabular}{lccccccccc}
\hline\hline
 & & \multicolumn{4}{c}{$M=1.4M_{\odot}$} & \multicolumn{4}{c}{$M=1.8M_{\odot}$} \\
\cmidrule(r){2-6} \cmidrule(r){7-10}
Model & $L_0$ &  $T_{\text{core}}$ & $T_{\rm{t}}$ & $\log Q_{\rm{imp}}$ & $R$ & $T_{\text{core}}$ & $T_{\rm{t}}$ & $\log Q_{\rm{imp}}$ & $R$ \\
 & (MeV) & ($10^7$ K) & ($10^8$ K) &  & (km) & ($10^7$ K) & ($10^8$ K) &  & (km) \\
\hline
 RMF
 & 40 & $4.28^{+0.19}_{-0.21}$ & $2.32^{+0.11}_{-0.11}$ & $0.52^{+0.08}_{-0.09}$ & 11.55 & $4.20^{+0.19}_{-0.21}$ & $2.80^{+0.16}_{-0.15}$ & $0.84^{+0.06}_{-0.07}$ & 11.10 \\
 & 50 & $4.29^{+0.20}_{-0.20}$ & $2.33^{+0.12}_{-0.12}$ & $0.57^{+0.08}_{-0.09}$ & 11.61 & $4.23^{+0.18}_{-0.19}$ & $2.83^{+0.16}_{-0.16}$ & $0.89^{+0.06}_{-0.06}$ & 11.12 \\
 & 60 & $4.32^{+0.19}_{-0.20}$ & $2.34^{+0.13}_{-0.12}$ & $0.63^{+0.07}_{-0.08}$ & 11.71 & $4.24^{+0.19}_{-0.18}$ & $2.82^{+0.19}_{-0.17}$ & $0.94^{+0.05}_{-0.06}$ & 11.15 \\
 & 70 & $4.35^{+0.18}_{-0.19}$ & $2.36^{+0.13}_{-0.12}$ & $0.67^{+0.07}_{-0.07}$ & 11.89 & $4.26^{+0.17}_{-0.17}$ & $2.81^{+0.17}_{-0.15}$ & $0.99^{+0.05}_{-0.05}$ & 11.23 \\
 & 80 & $4.46^{+0.18}_{-0.19}$ & $2.36^{+0.12}_{-0.12}$ & $0.69^{+0.06}_{-0.07}$ & 12.23 & $4.32^{+0.18}_{-0.17}$ & $2.75^{+0.17}_{-0.16}$ & $1.02^{+0.05}_{-0.05}$ & 11.50 \\
\hline
 QMF 
 & 40 & $4.32^{+0.20}_{-0.20}$ & $2.31^{+0.12}_{-0.12}$ & $0.49^{+0.08}_{-0.10}$ & 11.70 & $4.22^{+0.21}_{-0.20}$ & $2.72^{+0.15}_{-0.14}$ & $0.80^{+0.06}_{-0.07}$ & 11.48 \\
 & 50 & $4.33^{+0.20}_{-0.21}$ & $2.33^{+0.12}_{-0.12}$ & $0.55^{+0.08}_{-0.10}$ & 11.78 & $4.25^{+0.20}_{-0.20}$ & $2.75^{+0.15}_{-0.15}$ & $0.84^{+0.06}_{-0.07}$ & 11.51 \\
 & 60 & $4.34^{+0.19}_{-0.19}$ & $2.34^{+0.13}_{-0.12}$ & $0.61^{+0.07}_{-0.09}$ & 11.87 & $4.27^{+0.19}_{-0.18}$ & $2.75^{+0.15}_{-0.15}$ & $0.89^{+0.05}_{-0.06}$ & 11.56 \\
 & 70 & $4.40^{+0.19}_{-0.19}$ & $2.35^{+0.12}_{-0.12}$ & $0.65^{+0.07}_{-0.08}$ & 12.04 & $4.33^{+0.19}_{-0.19}$ & $2.74^{+0.17}_{-0.16}$ & $0.93^{+0.05}_{-0.06}$ & 11.67 \\
 & 80 & $4.47^{+0.20}_{-0.19}$ & $2.35^{+0.12}_{-0.12}$ & $0.68^{+0.06}_{-0.08}$ & 12.36 & $4.37^{+0.17}_{-0.18}$ & $2.71^{+0.16}_{-0.16}$ & $0.97^{+0.05}_{-0.06}$ & 11.94 \\
\hline\hline
\end{tabular}
\end{table*}

Different stages of crust cooling reflect the microscopic processes at different depths within the crust; longer cooling times correspond to greater depths~\cite{2009ApJ...698.1020B}. 
After the system reaches the quiescent phase, the core temperature can be inferred from the cooling curve, as $T_{\rm b}$ asymptotically approaches the core temperature, i.e., $T_{\rm b}\sim T_{\rm core}$. 
Since the observed temperature scales as $T_{\rm{eff}}^{\infty}\propto g^{1/4} / (1 + z)T_{\rm b}^{5.5}$, a larger stellar radius—which lowers the surface gravity $g$—leads to an increased inferred core temperature. Consequently, the QMF model, which predicts a larger radius, and effective interactions with higher $L_0$, both yield higher fitted core temperatures, as summarized in Table~\ref{tab:shorttermcooling}.
In addition to  to the symmetry energy slope values $L_0=40,~60$, and 80 MeV, we also included calculations for $L_0=50$ and 70 MeV in both the QMF and RMF models.
The impurity parameter $Q_{\rm imp}$ affects the thermal relaxation in transients. 
Compared to the RMF model, the QMF model predicts larger values of $A_{\rm{ion}} $ and $Z$, increasing the phonon-specific heat and reducing thermal conductivity via electron-phonon scattering, which slows thermal relaxation. 
To reproduce the observed thermal relaxation timescale, $Q_{\rm{imp}}$ must be reduced to weaken the electron-impurity scattering and offset the slowing of thermal relaxation. A similar trend is observed for the dependence on $L_0$: as $L_0$ increases, a higher $Q_{\rm{imp}}$ is needed. 
Table~\ref{tab:shorttermcooling} presents the fitting results for NSs with masses of $1.4\,M_{\odot}$ and $1.8\,M_{\odot}$.
We find that the qualitative dependencies of $T_{\rm core}$ and $Q_{\rm imp}$ on the model and $L_0$ remain unchanged with mass.
The parameter $T_{\rm t}$ is included to account for shallow heating effects~\cite{2024PhRvL.132r1001A}.The QMF model predicts a slightly lower $T_{\rm t}$ for canonical-mass NSs, and a significantly lower value for massive ones. 
For both canonical and massive stars, we observe a non-monotonic dependence of
$T_{\rm t}$ with $L_0$.
A more refined fitting would require treating both the strength and depth of the shallow heating layer as free parameters, 
which we leave for a detailed study in future work. 

\begin{figure}
{\centering
{\includegraphics[width=1.0\linewidth]{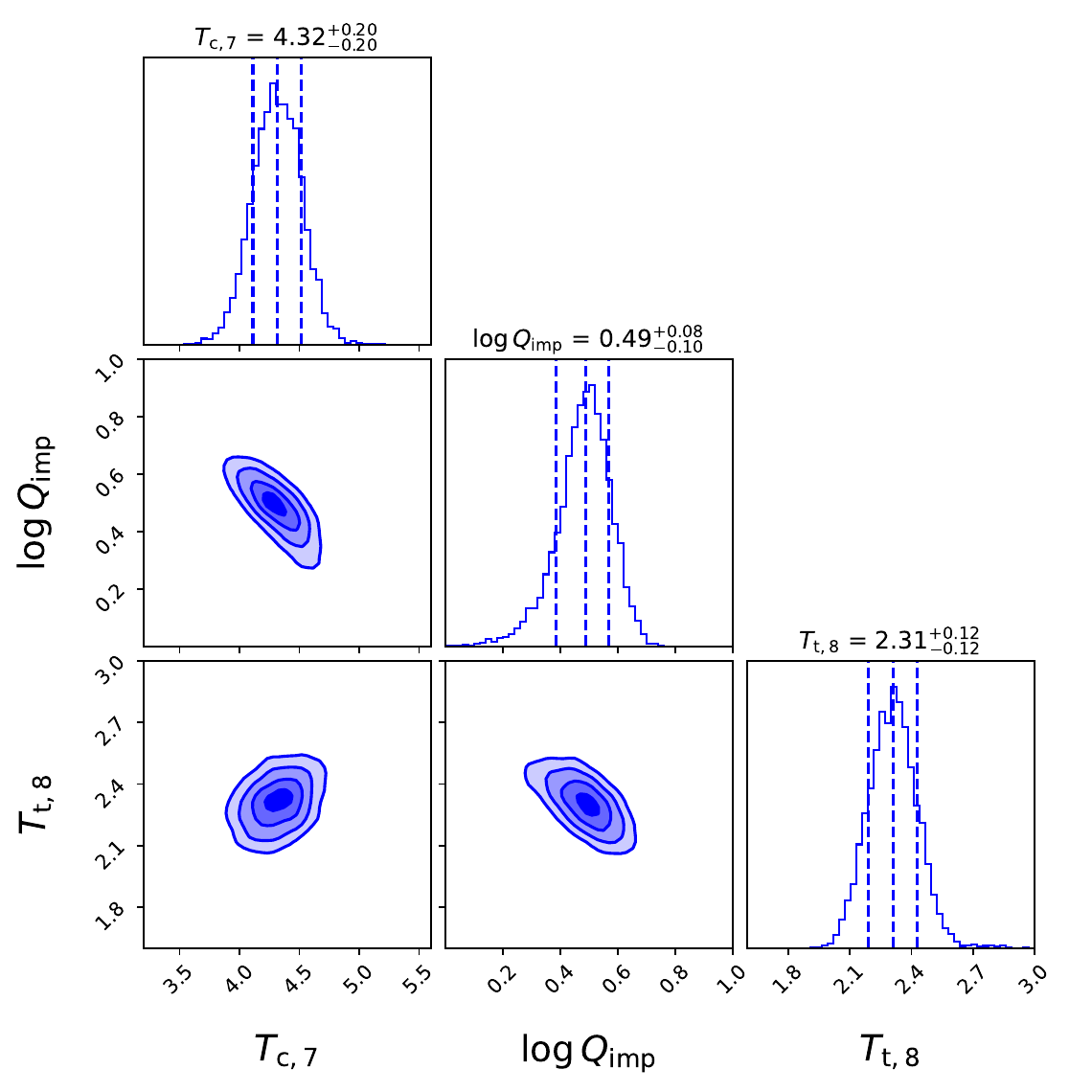}}}
\caption{Posterior distributions of $T_{\text{c},7}$, $\log Q_{\text{imp}}$,  and $T_{\text{t},8}$ for the crust cooling simulation of KS 1731-260. The effective interaction is the QMF model with $L_0=40$ MeV.  The NS mass is assumed as $1.4\,M_{\odot}$. $T_{\text{c},7}$ is $T_{\text{core}}$ in the unit of $10^7$ K, and $T_{\text{t},8}$ is $T_{\text{t}}$ in the unit of $10^8$ K.
}.\label{fig:corner}
\end{figure}

Assuming the mass of KS 1731-260 as $1.4\,M_{\odot}$ and then taking the QMF model with $L_0$ = 40 MeV as an example, we present the posterior distributions of the the fitting parameters—$T_{\mathrm{core}}$, $T_{\rm t}$, and $Q_{\rm{imp}}$—in Fig. \ref{fig:corner}. We find relatively weak degeneracies among them. Both the temperature $T_{\rm{t}}$ and $T_{\text{core}}$ exhibit anti-correlations with $Q_{\rm{imp}}$. For a fixed $T_{\rm{core}}$, the core neutrino emissivity remains unchanged, the higher $T_{\rm{t}}$ requires a lower $Q_{\rm{imp}}$ to accelerate cooling and match the observed thermal relaxation timescale. For a fixed $T_{\rm{t}}$, the higher $Q_{\rm{imp}}$ enhances the core neutrino emissivity. To maintain thermal equilibrium between the crust and the core after cooling, the inward heat flux through the crust must increase, which requires a lower $Q_{\rm{imp}}$ to increase thermal conductivity. Weaker correlation is observed between $T_{\rm{t}}$ and $T_{\rm{core}}$. This is reasonable because $T_{\rm{t}}$ reflects primarily the thermal properties of the crust at the bottom of the envelope during the early cooling stages and is more strongly related to outburst and shallow heating mechanisms.

\begin{figure}
{\centering
{\includegraphics[width=1.0\linewidth]{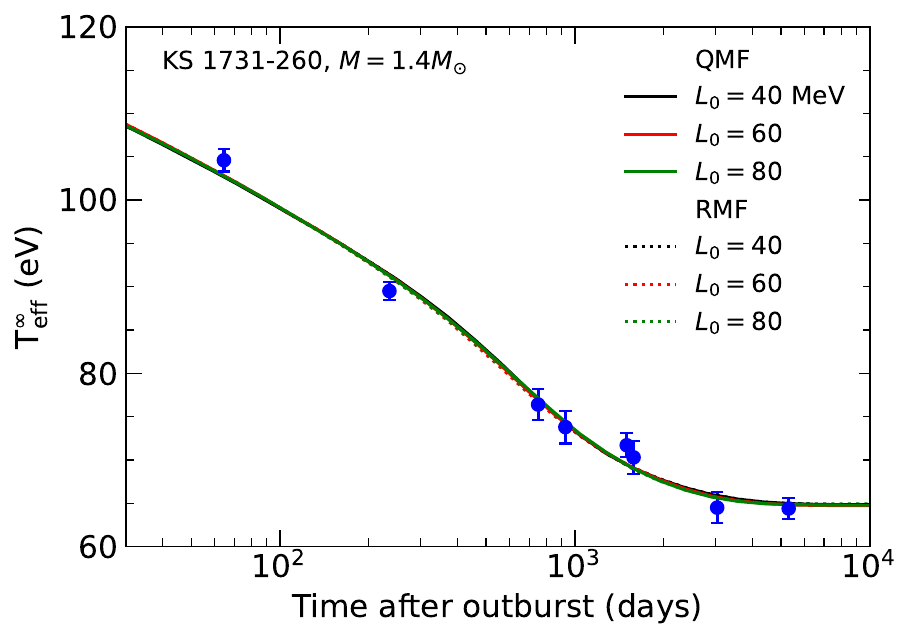}}}
\caption{Light curves for KS 1731-260 based on the QMF and RMF models with different $L_0$ assuming a canonical NS mass of $1.4 M_{\odot}$. Observational data are taken from Ref. \cite{Merritt2016_ApJ833-186}. 
}.\label{fig:crustcooling}
\end{figure}

Crustal thermal relaxation in X-ray transients following accretion cessation provides a valuable probe of crustal properties.
Fig. \ref{fig:crustcooling} shows the best-fit cooling curves for KS 1731-260, assuming a NS mass of $1.4 M_{\odot}$. 
All fitted cooling curves appear nearly degenerate; however, slight differences in the three fitting parameters are summarized in Table~\ref{tab:shorttermcooling}.

Using the fitted core temperature of KS 1731-260, $T_{\mathrm{core}}\approx4.4\times10^7$ K, we proceed to investigate the r-mode oscillations of NSs.

\subsection{Viscosity and r-mode instability }\label{sec:result_viscosity}

\begin{figure}
{\centering
{\includegraphics[width=1.0\linewidth]{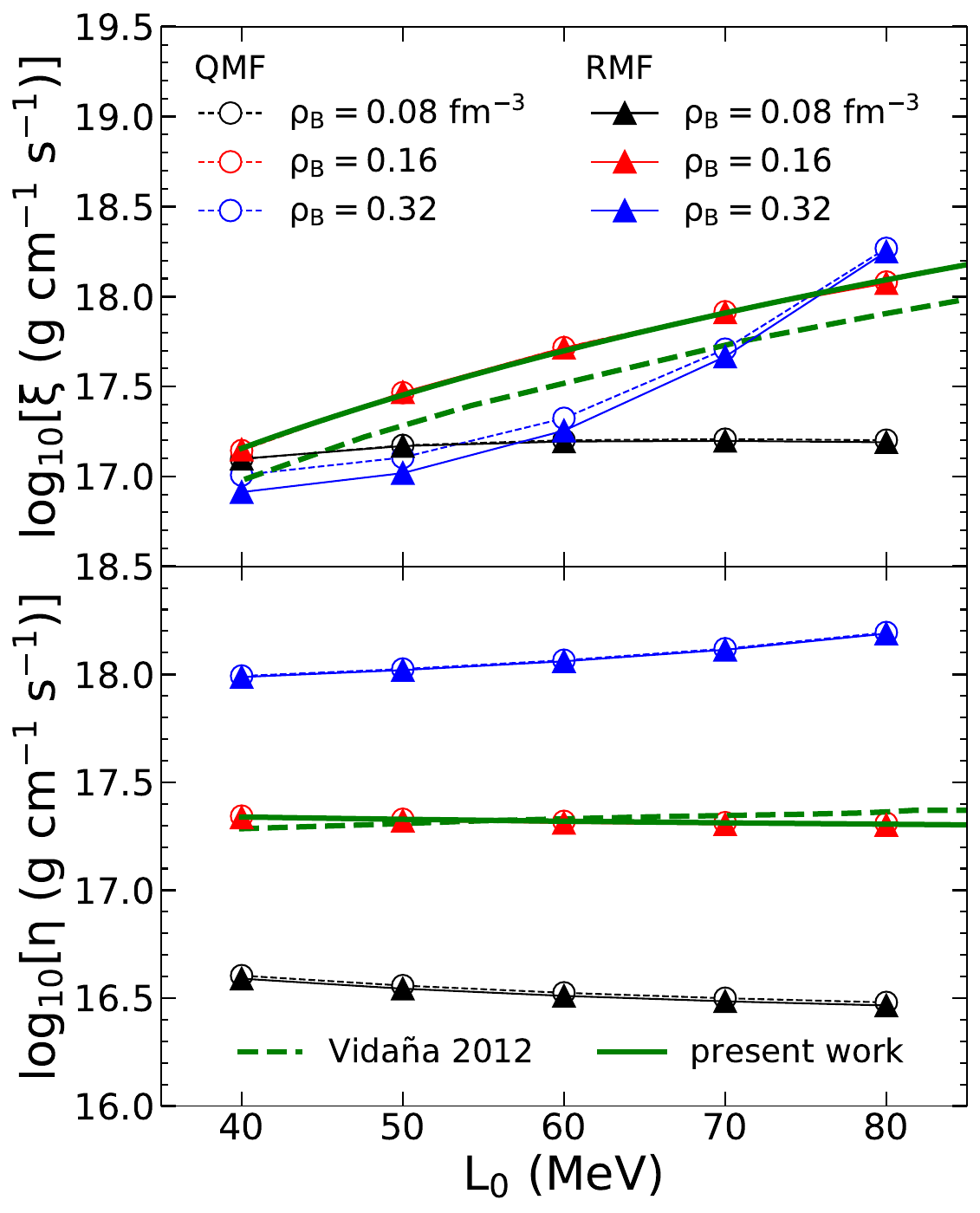}
}
}
\caption{Bulk (upper panel) and shear (lower panel) viscosities as functions of symmetry energy slope $L_0$ at several densities. The green solid lines represent the power-law fits $\mathrm{log_{10}}\xi=12.9285L_0^{0.0767}$ and $\mathrm{log_{10}}\eta=17.5192L_0^{-0.0028}$, while the dashed lines correspond to the results from Ref.~\citep{2012PhRvC..85d5808V}.
}\label{fig:viscosity}
\end{figure}

\begin{figure}
{\centering
{\includegraphics[width=1.0\linewidth]{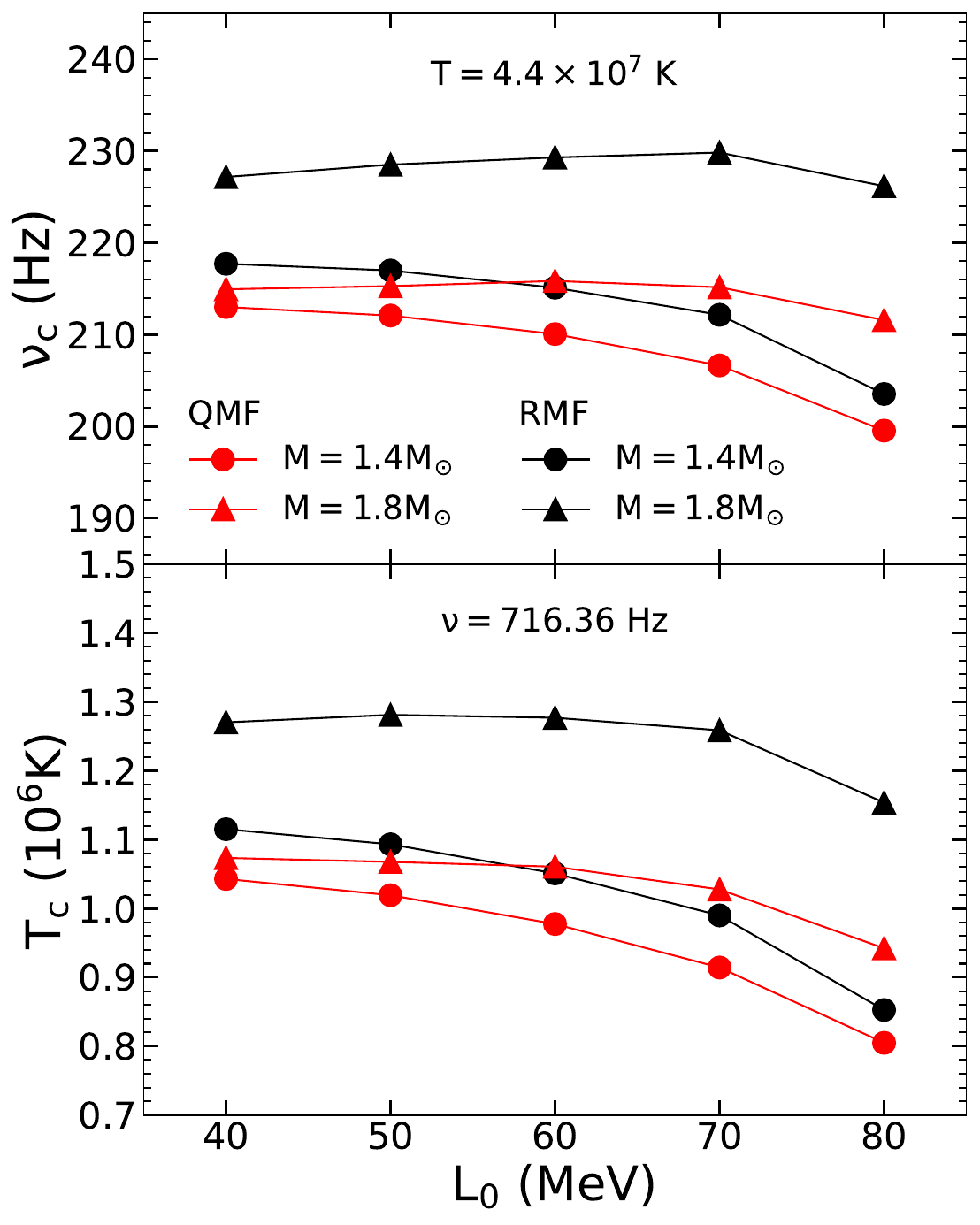}
}
}
\caption{Critical frequency $\Omega_c$ (upper panel) and critical temperature $T_c$ (lower panel) as a function of the symmetry energy slope $L_0$ for a $1.4M_{\odot}$ and $1.8M_{\odot}$ NS.}\label{fig:OCTC}
\end{figure}

According to Eqs.~(\ref{eq:DBulk})–(\ref{eq:sheare}), both viscosities depend sensitively on the properties of nuclear matter.
Figure~\ref{fig:viscosity} shows the dependence of bulk and shear viscosities on the symmetry energy slope parameter $L_0$ for $T=10^9\ \mathrm{K}$ and an r-mode frequency of $\omega=10^4\ \mathrm{s^{-1}}$, as calculated using the QMF and RMF models. 
Overall, the viscosities predicted by the two models are quite similar, with a noticeable difference in bulk viscosity appearing only at high densities ($\rho_B = 0.32~\text{fm}^{-3}$). This difference is attributed to the divergent behavior of the symmetry energy derivatives with respect to density in the two models at such high densities.
The dependence of $\xi$ and $\eta$ on $L_0$ can be approximated by simple power laws of the form $\mathrm{log_{10}}\xi=12.9285L_0^{0.0767}$ and $\mathrm{log_{10}}\eta=17.5192L_0^{-0.0028}$ at saturation density, as indicated by the solid lines in the figure. These results are shown in comparison with the findings of Ref.~\citep{2012PhRvC..85d5808V}, represented by dashed lines.

The r-mode instability window is sensitive to the nuclear matter EoS and depends on the symmetry energy slope parameter $L_0$~\citep{2012PhRvC..85d5808V,2021ApJ...910...62Z,2011PrPNP..66..519N}.
To explore this dependence, we investigate the critical frequency $\nu_c$ at a fixed temperature $T=4.4\times10^7\ \mathrm{K}$ (the fitted average core temperature of KS 1731-260), and the critical temperature $T_c$ at a fixed frequency $\nu=716.36\ \mathrm{Hz}$ (the highest observed spin frequency~\citep{2006Sci...311.1901H}). The results are shown in Fig.~\ref{fig:OCTC} for both QMF and RMF models, considering neutron stars with masses of $1.4M_{\odot}$ and $1.8M_{\odot}$. 
From Fig.~\ref{fig:MR}(a), we see that a larger symmetry energy slope $L_0$ leads to an increase in the NS radius. Correspondingly, both the critical frequency $\nu_c$ and the critical temperature $T_c$ decrease with increasing $L_0$, indicating a broader r-mode instability window at higher $L_0$.
This trend is observed consistently in both the QMF and RMF models, in contrast to the findings reported in Ref.~\citep{2012PhRvC..85d5808V} based on nonunified EoSs. 
The QMF model, which generally predicts larger radii for a given mass (see Fig.~\ref{fig:MR}), results in slightly lower values of $\nu_c$ and $T_c$, with the differences becoming more noticeable in more massive stars ($1.8M_{\odot}$).  
For $1.4M_{\odot}$ stars, the radius shows a stronger dependence on $L_0$, making both $\nu_c$ and $T_c$ more sensitive to variations in the symmetry energy slope. 
This implies a more pronounced $L_0$-dependence of the r-mode instability window in lower-mass NSs.
Furthermore, the values of $\nu_c$ and $T_c$ are higher for $1.8M_{\odot}$ stars, suggesting that the r-mode instability window narrows with increasing stellar mass.
In the RMF model, the differences in $T_c$ and $\nu_c$ between $1.4 M_{\odot}$ and $1.8 M_{\odot}$ NSs are even more significant, indicating a stronger variation in the r-mode instability window with mass compared to the QMF model.

\section{Summary}\label{sec:summary}
In this study, we construct several unified EoSs from the quark level in the framework of the QMF.
These EoSs are then used to study the global properties and dynamical behavior of NSs, in comparison to the results with the corresponding RMF EoSs. We evaluate the performance of the unified QMF EoSs in describing the $M$-$R$ relation, cooling of the isolated NS and X-ray transient, the viscosity and r-mode instability, analyzing especially the model and $L_0$ dependence of these properties.

In the crust, the slightly stronger $\sigma$ field in the QMF model leads to larger $A_{\rm{cell}}$, $A_{\rm{ion}}$, and $Z$ through the derivative terms of the equation of motion for the $\sigma$ meson in inhomogeneous matter. However, the density of free neutrons in the inner crust shows no significant model dependence due to the enlarged WS cell size in the QMF model, allowing us to safely adopt the same neutron $^1S_0$ pairing gap in both QMF and RMF models following BCS description. 
The proton fraction in the crust exhibits negligible model dependence.

Differences in crust composition do not significantly influence the EoS in the low-density range. The QMF model predicts higher pressures above 0.4 fm$^{-3}$. Consequently, the $M$-$R$ relations of NSs are dominated by $L_0$ for low central densities or light NS and by the model for high central densities or massive NS. 
For the cooling of isolated NSs, the cooling curves are insensitive to both the model and $L_0$ in the absence of the dUrca process. However, the QMF model predicts a longer thermal relaxation time in an NS undergoing rapid cooling driven by the dUrca process. Both models can reproduce cooling curves consistent with observations of KS 1731-260 during the crustal cooling, despite slight model- and $L_0$-dependent variations in stellar parameters. 
For larger symmetry energy slopes $L_0$, the r-mode instability window becomes wider. For more massive stars, the r-mode instability window is smaller. 

We emphasize that our analysis of the global properties and dynamical behavior is based on a consistent picture accompanying the unified EoS.
This combined analysis provides a comprehensive view of the interplay between dense matter microphysics and observable NS phenomena, contributing to ongoing efforts~\cite{2022EPJWC.26004001L} to constrain the EoS and the transport properties of dense nuclear matter.
Accumulating observational signatures~\cite{lattimer2007neutron,ascenzi2024neutron,2025arXiv250608104L} of these dynamic processes, including surface temperature evolution, timing irregularities, quasi-periodic oscillations, and gravitational wave emission, provide multiple avenues to probe the obscure composition of the stellar interior. 
From a theoretical standpoint, a more self-consistent microscopic description is essential for reliably predicting the dynamical properties of NSs.
The theoretical limitations in this study—such as the absence of self-consistent calculations for superfluid critical temperatures in the core and the neglect of finite-size effects of nuclei on crustal cooling—will be addressed in future work.

\acknowledgments
The work is supported by the National SKA Program of China (No.~2020SKA0120300), the National Natural Science Foundation of China (grant Nos. 12273028, 12494572).
Z.-Q. Miao wishes to acknowledge the support of the China national postdoctoral program for innovation talents (No.~BX20240223) and the China Postdoctoral Science Foundation funded project (No.~2024M761948).

\section*{Data Availability}
The data that support the findings of this article are openly available \cite{2017PhRvL.119p1101A,2024ApJ...961...62V,2024ApJ...974..294S,Beloin2018_PRC97-015804,Merritt2016_ApJ833-186}. 

\bibliography{unified}
\bibliographystyle{apsrev4-1}

\end{document}